\documentclass{aa}
\usepackage{txfonts}
\usepackage{graphicx}
\usepackage{natbib}
\usepackage{amstext,cases}
\bibpunct{(}{)}{;}{a}{}{,}
\newcommand{\grsim}{\mathrel{\hbox{\rlap{\hbox{\lower4pt\hbox{$\sim$}}}\hbox{$>$}}}}

\begin{document}

\title{Confidence limits of evolutionary synthesis models}
\subtitle{IV. Moving forward to a probabilistic formulation.}

\author{M. Cervi\~no \inst{1},
        V. Luridiana\inst{1},
\and
        N. Cervi\~no-Luridiana\inst{2}}

\institute{Instituto de Astrof\'\i sica de Andaluc\'\i a (CSIC), Camino bajo
        de Hu\'etor 50, Apdo. 3004, Granada 18080, Spain
\and E.I. Bel\'en,  Camino bajo de Hu\'etor s/n, Granada 18008, Spain\\
}

\offprints{M. Cervi\~no, V. Luridiana; \email{mcs@iaa.es, vale@iaa.es}}

\date{Received 22 April 2005; Accepted 30 December 2005}

\abstract
 % context heading (optional)
{Synthesis models predict the integrated properties of stellar populations.
Several problems exist in this field, mostly related to the fact that integrated properties are distributed. To date, this aspect has been either ignored (as in standard synthesis models, which are inherently deterministic) or interpreted phenomenologically (as in Monte Carlo simulations, which describe distributed properties rather than explain them).}
 % aims heading (mandatory)
{This paper presents a method of population synthesis that accounts for the distributed nature of stellar properties.}
% methods heading (mandatory)
{We approach population synthesis as a problem in probability theory, in which stellar luminosities are 
random variables extracted from the stellar luminosity distribution function (sLDF).}
% results heading (mandatory)
{With standard distribution theory, we derive the population LDF (pLDF) for clusters of any size from the sLDF, obtaining the scale relations that link the sLDF to the pLDF. We recover the predictions of standard synthesis models, which are shown to compute the mean of the luminosity function.
We provide diagnostic diagrams and a simplified recipe for testing the statistical richness of observed clusters, thereby assessing whether standard synthesis models can be safely used or a statistical treatment is mandatory.
We also recover the predictions of Monte Carlo simulations, with the additional bonus of being able to interpret them in mathematical and physical terms. 
We give examples of problems that can be addressed through our probabilistic formalism: calibrating the SBF method, determining the luminosity function of globular clusters, comparing different isochrone sets, tracing the sLDF by means of resolved data, including fuzzy stellar properties in population synthesis, among others.
Additionally, the algorithmic nature of our method makes it suitable for developing analysis tools for the Virtual Observatory.}
%conclusions heading (optional), leave it empty if necessary
{Though still under development, ours is a powerful
approach to population synthesis. In an era of resolved observations and pipelined analyses of large surveys, this paper is offered as a signpost in the field of stellar populations.}


\titlerunning{Probabilistic synthesis models}
\maketitle

\section{Introduction and motivation}

The study of stellar populations is one of the
most fecund topics in today's astronomy. Understanding
the properties of stellar populations is a key element
in the solution of a host of fundamental problems, 
such as the calibration of distances to extragalactic objects,
the age determination of clusters and galaxies through color fitting,
the characterization of the star-formation history of 
composite populations, the modeling of the chemical evolution of galaxies, 
and several more.
When tackling these problems, the interaction between theory and observations
goes both ways: one may want to predict 
the properties of a stellar population with given properties,
or to recover the basic properties of an observed population
from the available observables. In either case, to accomplish the task
one needs feasible theoretical models that serve as reliable diagnostic tools.
 
A traditional approach to building such diagnostic tools 
has been the computation of synthesis models. 
Synthesis models allow one to predict the features and the evolution of
stellar populations in a highly structured way, one that is 
apt for routine analysis and quantitative assessments.
For example, synthesis models can be used, in combination with
other methods, for the determination of stellar population properties of 
large samples of galaxies in a reasonable time;
e.g. the analysis of 50362 galaxies of the Sloan Digital Sky Survey 
based on their integrated properties performed by \cite{CFetal05}.

However, in recent years there has been a growing awareness 
that synthesis modeling also suffers from severe limitations.
In some cases, we know {\it a priori} that standard models cannot be applied
to a given system, because the properties of the system fall
outside the hypothesis space of the models; this is the case,
for example, of undersampled populations \citep{Chietal88,CVG03}. In other cases,
we observe a mismatch between the properties of the system
inferred from observations of their individual components and 
those derived from their integrated properties analyzed with synthesis models:
e.g. the discrepancy between the age determined from the color-magnitude diagram and the spectroscopic age in NGC~588 \citep{Jametal04}, or the IMF slope inferred by \cite{P05} in undersampled giant H{\sc~ii} regions. 
In these cases, we are facing
a crisis in the explanatory power of synthesis models.

Previous attempts to solve this crisis and to understand the 
limitations of synthesis models have repeatedly pointed at the necessity
of including statistical considerations in the analysis.
According to this view, the predictive power breaks down because 
the traditional synthesis model is essentially a deterministic
tool, whereas the nature of the problem is inherently stochastic.
The clearest example of stochasticity is the mass spectrum of 
stellar populations, in which fluctuations (possibly random, although not necessarily so) in the number of stars of each type 
appear around the mean expected number.
Until now, the efforts to take stochasticity into account in the modeling
of stellar populations have moved in essentially two directions:
the use of Monte Carlo techniques \citep[e.g.][among others]{SF97,CLC00,Brutuc,Gituc,CRBC03} and  the reinterpretation of
traditional models in statistical terms \cite[e.g.][]{Buzz89,CVGLMH02,GLB04}. Both methods have proved able
to solve some of the problems, or at least point toward possible solutions.
However, they suffer from practical difficulties.
Monte Carlo simulations are expensive 
(in terms of disk space, CPU time, and 
human time required to analyze the results),
while, to date, the statistical reinterpretation of standard models has only 
served
to establish limits to the use of synthesis models for clusters with moderately undersampled populations \citep{CL04,Cetal03}, i.e. clusters with total initial masses on the order of $10^5$ M$_\odot$. 

This limitation brings about a serious impasse for the study of stellar populations by means of their integrated light, since the properties of clusters with lower masses cannot be reliably obtained. This class includes, for example, all of the clusters in our Galaxy (including globular clusters), clusters in the 
Large Magellanic Cloud (LMC),
as well as many clusters in external galaxies.
For example, 
many of the clusters in the `Antennae' studied by \cite{ZF99}
have masses lower than this limit: 
in Sect.~\ref{sub:gclf} we will show that 
explicit consideration of stochastic effects could alter the
conclusions that are drawn on the cluster luminosity function based on these clusters.
Furthermore, these limitations will become even more dramatic with the development of Virtual Observatory (VO) technologies, which will make the automatic analysis of large amounts of data possible.  It is therefore mandatory to 
adapt evolutionary synthesis models to the present needs, 
so that they can
be applied to observations.

In the present paper we introduce a theoretical formalism for the 
probabilistic 
analysis of single stellar populations (SSPs). Our formalism yields results that are as accurate as those of large Monte Carlo simulations, but it bypasses the need to perform these simulations: that is, the method is both accurate and economic. By means of our formalism, synthesis models can be applied to clusters of any size and the confidence intervals of the results can be evaluated easily. This makes it possible to estimate the relative weights of different bands for the realistic application of goodness-of-fit criteria like the $\chi^2$ test. Finally, the algorithmic nature of our method makes it feasible for implementation in the VO environment. 

This paper is the fourth in a series \citep{CLC00,CVGLMH02,Cetal01} dealing with the statistical analysis of stellar populations. Although it only deals with SSPs, a fifth paper, in preparation, will be devoted to the extension of this formalism to any star formation history scenario.
Finally, in \cite{CL05} we give an extensive review of the uncertainties affecting the results of synthesis models.
In addition to these papers,
we also recommend the work by \cite{GGS04}: {\it Statistical properties of the combined emission of a population of discrete sources: astrophysical implications}. This paper, although not directly focused on synthesis models, suggests an alternative point of view of the problem. In some aspects, it has inspired the present paper.  

In this work we restate the problem of 
stellar population synthesis from a new
perspective, the one of luminosity distributions, which is
a powerful and elegant way to understand stellar populations.
We provide several examples of applications to illustrate such power 
and indicate potential areas of future development.
The starting point of the new formalism is the definition of the stellar (i.e. individual) 
luminosity distribution function (LDF) and of its relation to 
the central problem of population synthesis (Sect.~\ref{sect:overview}). 
Sect.~\ref{sect:basic} describes the two main variants of synthesis models
by means of which the problem has traditionally been tackled:
Monte Carlo and standard models.
Next, we take the reader 
on a journey through the no man's land
of population synthesis' pitfalls (Sect.~\ref{sect:pitfalls}),
and show how these are faced by
existing techniques (Sect.~\ref{sect:algorithms}). 
This account is necessary
to introduce, in Sect.~\ref{sect:probabilistic}, our suggested solution 
and show its comparative power.
Finally, in Sect.~\ref{sect:application}, 
we give a few examples of applications
of our formalism to current problems. Future developments
of the present work are described in Sect.~\ref{sect:future},
and our main conclusions
are summarized in Sect.~\ref{sect:conclusions}.

\section{Overview of the problem}\label{sect:overview}

This section starts by introducing a few basic definitions. An exhaustive summary of the notation used and its rationale is offered in Appendix~\ref{app:notation}.

The general problem approached by synthesis models is the computation
of the luminosity\footnote{Throughout the paper, `luminosity' 
will be a generic
label for the stellar emission in any given band.} $L_{\mathrm{tot}}$ emitted by an ensemble of 
$N_{\mathrm{tot}}$ sources -- a stellar population.
From a theoretical point of view, this problem 
can be characterized in three basic ways. 
 
If the luminosities $l_i$ of the individual sources are known, 
the total luminosity $L_{\mathrm{tot}}$ 
is obtained trivially as the sum of all the $l_i$ values:

\begin{equation}
L_{\mathrm{tot}} = \sum_{i=1}^{N{_\mathrm{tot}}}  \, l_i.
\label{eq:Ltot_sum}
\end{equation}

\noindent This circumstance is not very frequent. Its most common 
examples are Monte Carlo synthetic clusters (Sect.~\ref{sect:basic}) in the theoretical domain
and resolved stellar populations in the observational one.

In the more usual situation in which the luminosities of individual 
objects 
cannot be specified on an individual basis,
a different approach must be adopted:
in this case a luminosity distribution function (LDF) $\varphi_\mathrm{L}(\ell)$
is 
assumed that describes the 
distribution of luminosity 
values in a generic ensemble. Traditionally, the LDF
has been seen as the asymptotic limit 
of a differential count of stars:

\begin{equation}
\varphi_\mathrm{L}(\ell) = \lim_{N_{\mathrm{tot}} \rightarrow \infty}\Bigl({{1\over{N_{\mathrm{tot}}}}\lim_{\Delta \ell \rightarrow 0}\frac{\Delta N}{\Delta \ell}}\Bigr) \, ,
\label{eq:LDFtrad}
\end{equation}

\noindent where $\Delta N$ is the number of stars in a luminosity bin of width $\Delta \ell$,
and $N_{\mathrm{tot}}$ the total number of stars. The integral of the LDF is normalized to 1:

\begin{equation}
\int_{0}^{\infty} \varphi_\mathrm{L}(\ell)d\ell = 1.
\label{eq:normLDFtrad}
\end{equation}

\noindent The integrated luminosity of an ensemble
is traditionally obtained by means of the expression:

\begin{equation}
L_{\mathrm{tot}} = N_{\mathrm{tot}} \int_{0}^{\infty} \ell \, \varphi_\mathrm{L}(\ell)d\ell.
\label{eq:Ltot_int}
\end{equation}

The bottom line of the present work is that this approach 
is conceptually wrong and operationally sterile.
The crucial point where we part company with this approach is
the definition and interpretation of the LDF: 
to us, $\varphi_\mathrm{L}(\ell)$ is a  {\it probability density function} 
(PDF) from which the luminosity of an actual system is drawn.
If the system is an individual star, its PDF 
is the {\it stellar} LDF (sLDF).
If it is a stellar population,
its PDF is the {\it population} LDF (pLDF).
Hereafter, we will indicate 
the sLDF with the symbol $\varphi_\mathrm{L}(\ell)$
and the pLDF with the symbol $\varphi_\mathrm{L_{tot}}({\cal L})$.

To avoid confusion, note that the pLDF is not the same as the
cluster or galaxy luminosity function (LF) as commonly defined, because 
galaxy LFs include the effect of a spread in ages and number of stars 
(or, equivalently, ages and mass), whereas the pLDF defined in this paper
represents the luminosity distribution of ensembles of stars with the 
same physical parameters (age and total number of stars). 
 
In this paper, we are concerned with the properties of the sLDF 
and its relation to the pLDF. In particular, we will show that
the total luminosity of a stellar population, ${\cal L}$, is a distributed quantity,
whose {\it mean} value $M_1'$ is given by:

\begin{equation}
M_1'  \equiv \langle {\cal L}\rangle = N_{\mathrm{tot}} \int_{0}^{\infty} \ell \, \varphi_\mathrm{L}(\ell)d\ell.
\label{eq:Ltot_mean}
\end{equation}

\noindent The integral on the right-hand side of this equation is the mean value of the sLDF, $\mu_1'$, that is:

\begin{equation}
M_1' = N_{\mathrm{tot}} \, \mu_1'.
\label{eq:Mu_Ntotmu}
\end{equation}

\noindent 

Although Eqs.~\ref{eq:Ltot_int} and \ref{eq:Ltot_mean} yield similar expressions,
it is important to distinguish between the two approaches.
The historical origin of our 
sLDF
lies in star counts,
but it is a step further from them, in the same sense
that the frequentist and the objectivist definitions of probability differ:
the frequentist definition depends on the realization of trials, 
while the objectivist definition assumes that the probability properties
are built-in.
The implications of either approach
will be discussed at length throughout the paper.
For the time being, note that Eq.~\ref{eq:Ltot_sum} involves a discrete sum,
while Eqs.~\ref{eq:Ltot_int} and \ref{eq:Ltot_mean} involve an integral,
whose numerical solution requires binning the independent variable.
In Sect.~\ref{sub:binning} we discuss the consequences of 
binning distribution functions for numerical applications.

The main goal of synthesis models is to obtain the 
luminosity of a model stellar population, either by direct count 
(Eq.~\ref{eq:Ltot_sum}) or by an integral including the 
sLDF
(Eqs.~\ref{eq:Ltot_int} and \ref{eq:Ltot_mean}). In the following, we will see how this 
can be carried out in practice, under the assumption that
there has been a star-forming episode in which all the stars have
formed simultaneously; this is the scenario assumed by 
SSP models.

Because stellar luminosities evolve with time, 
the sLDF is a function of the star's age.
Since the luminosity evolution of a star depends on its initial mass,
we can express the time dependence of the sLDF explicitly
by writing the sLDF as a function of two other functions: 
the isochrone $\ell(m;t)$ and 
the initial mass function (IMF) $\varphi_{\mathrm{M}}(m)$.
The isochrone $\ell(m;t)$ gives the luminosity of a star as 
a function of its initial mass $m$ at a given value of the age $t$.
The IMF gives 
the probability distribution of initial stellar masses. The IMF
has a status similar to that of the
sLDF,
in that it can be either interpreted as the result of a star count:

\begin{equation}
\varphi_\mathrm{M}(m) = \lim_{N_{\mathrm{tot}} \rightarrow \infty}\Bigl({{1\over{N_{\mathrm{tot}}}}\lim_{\Delta m \rightarrow 0}\frac{\Delta N}{\Delta m}}\Bigr) \, ,
\label{eq:IMFtrad}
\end{equation}

\noindent or, in probabilistic terms, as the probability density for a star of being
born with mass $m$: for reasons similar to those
discussed above, we support the second interpretation (which, by the
way, implies that the IMF should better be called Initial Mass 
{\it Probability Density} Function).
According to its definition, $\varphi_{\mathrm{M}}(m)$ fulfills
the normalization condition:

\begin{equation}
\int_{0}^{\infty} \varphi_\mathrm{M}(m)dm = 1.
\label{eq:IMF1}
\end{equation}

Summing up, the sLDF can be rewritten in terms of the IMF and the isochrone as follows:

\begin{equation}
\varphi_{\mathrm{L}}(\ell;t) = \varphi_\mathrm{M}(m) \times \biggl( \frac{d\ell(m;t)}{dm} \biggr)^{-1}.
\label{eq:ldf-imf}
\end{equation}

\noindent Eq.~\ref{eq:ldf-imf} can now be used to rewrite the mean value of the
sLDF
in terms of the isochrone and the IMF\footnote{The isochrone is not monotonic,
so that the integral limits of Eq.~\ref{eq:Ltot_mean} do not correspond to those of Eq. \ref{eq:muIMF}.}:

\begin{eqnarray}
\mu_1'(t) &=& \int_{m^\mathrm{low}}^{m^{\mathrm{up}}} \ell(m;t) \, \varphi_\mathrm{M}(m) \, \biggr(\frac{d\ell(m;t)}{dm}\biggl)^{-1} \,  \frac{d\ell(m;t)}{dm}\, dm  \nonumber = \\ 
               &= &  \int_{m^\mathrm{low}}^{m^{\mathrm{up}}} \ell(m;t) \, \varphi_\mathrm{M}(m)\, dm,
               \label{eq:muIMF}
\end{eqnarray}

\noindent where the integration variable in Eq. ~\ref{eq:Ltot_mean} has been changed from $\ell$ to $m$, and ${m^\mathrm{low}}$, ${m^{\mathrm{up}}}$ are the lower and 
upper mass limits respectively of the integration domain.

Solving this integral is the main task of stellar population synthesis modeling. In the following section we will describe the main types of synthesis models and how they perform this integral.

\section{Basic strategies}\label{sect:basic}

Once the physical problem is framed, we must translate it into 
actual computations, a task carried out by
evolutionary synthesis codes. These come in two basic flavors,
standard and Monte Carlo.  Standard simulations are models 
in which the initial mass mixture is analytically described
by the IMF,
whereas in Monte Carlo simulations the ensemble of stars is selected by random
sampling the mass of each star to be included in the population,
and the IMF is used as the weighting function.
Therefore, the mass distribution obtained with a standard
simulation is univocally determined by the population's parameters,
while in Monte Carlo simulations it is not; additionally,
the standard distribution is smoother than the Monte Carlo one.
Both methods, however, operate on a binned mass spectrum, due to the limited
resolution of numerical computation and to the discreteness of the
available stellar models. This fact has important consequences, which will
be discussed in Sect.~\ref{sub:binning}.

Using either of these two approaches, evolutionary synthesis codes
aim to characterize the
integrated emission properties of an ensemble of stars as a function of
its physical parameters, such as the age and number of the individual stars
of the ensemble\footnote{Strictly speaking, not all the
luminosity sources contributing to the integrated luminosity
of a stellar population need be stars, as they include, e.g.,
accretion disks or thermalization of kinetic energy. 
Accordingly, one should in principle use the more general expression 
`luminosity sources' rather than `stars' to avoid loss of generality.
This distinction, however, is not relevant for our scope here, 
since in our treatment any luminosity source can be reconducted 
to a star. Therefore, in the following we will
use `stars' and `luminosity sources' as synonyms.}. 
Standard codes perform this task by carrying out the integration
of Eq.~\ref{eq:muIMF}. As described in Sect.~\ref{sect:overview},
the result can be interpreted in two alternative ways:
either a deterministic one -- $\mu_1'$ is the
sum of the luminosities of all the stars included in the ensemble modeled,
normalized to one star --,
or a probabilistic one -- $\mu_1'$ is the mean value of 
the sLDF.
Although the two interpretations may seem close at first sight, 
they are fundamentally different, both from a conceptual
and from a practical perspective:
this point will be furthered in Sect.~\ref{sect:probabilistic}.
We call these two alternative interpretations of standard synthesis
models the {\it deterministic} and the {\it statistical} one. 
Note that these labels do not identify different classes of
models, but rather different interpretations within the same class of
models -- standard models. In practice, some codes do not explore
this interpretation, while others acknowledge the distributed 
nature of luminosity and
compute, in addition to the mean luminosity,
the variance of the distribution.
In either case, $\mu_1'$ 
can eventually be scaled
to the size of the ensemble
by multiplying by $N_{\mathrm{tot}}$;
this property will be formally demonstrated in Sect.~\ref{sect:s2pLDF}.

Note
that many codes do not use $\varphi_{\mathrm{M}}(m)$
in Eq.~\ref{eq:muIMF}, but rather a proportional function 
$\varphi'_{\mathrm{M}}(m) = \mathrm{const}\, \varphi_{\mathrm{M}}(m)$ normalized in such a way that:

\begin{equation}
\int_{0}^{\infty} m \, \varphi'_\mathrm{M}(m)dm = 1.
\label{eq:IMF2}
\end{equation}

\noindent Since $\int_{0}^{\infty} m \, \varphi_\mathrm{M}(m)dm$ is the mean
mass value $\langle m \rangle$ of the IMF, using $\varphi'_{\mathrm{M}}(m)$ instead
of $\varphi_{\mathrm{M}}(m)$ in
Eq.~\ref{eq:muIMF} yields the mean luminosity of one
star divided by $\langle m \rangle$: in this case, the mean total 
luminosity is found by multiplying by the total mass of the ensemble
$M_{\mathrm{tot}}$ instead of $N_{\mathrm{tot}}$.

As for Monte Carlo models, their task is essentially 
the computation of the sum in Eq.~\ref{eq:Ltot_sum}. 
Each time a simulation is run, a particular realization is drawn 
from the underlying distribution.
The result of the simulation is the integrated luminosity $\cal L$
of that particular realization.
If many Monte Carlo simulations are available for a fixed set of 
input parameters, an estimate for
$M_1'$ can be obtained as the mean of the $\cal L$ values.
This implies that the results of Monte Carlo simulations depend not only on the underlying luminosity distribution, but also on the number of simulations used to sample such distribution.
If the set of simulations is sufficiently large,
the Monte Carlo method has also the potential to provide
the actual distribution function of the 
luminosities of the ensemble.
Hence, an important drawback of the method is that it is intrinsically expensive, because the accuracy of the results increases with the number of simulations.

In summary, the goal of determining the luminosity of stellar populations
reduces to the computation of a sum (Eq.~\ref{eq:Ltot_sum})
or an integral (Eq.~\ref{eq:muIMF}). However straightforward this may
seem, this computation is hindered in the practice by several intrinsic 
features of the problem: these will be the topic of next section.

\section{Pitfalls in the handling of the LDF}\label{sect:pitfalls}

Our previous discussion has taken place on an abstract level.
In the practice of synthesis modeling it is necessary to
translate the concepts discussed above into specific prescriptions for the handling
of equations, and deal with the limitations imposed by the input ingredients,
which have a finite resolution in the parameter space.
In this section we will discuss a specific aspect of this task,
namely the difficulties inherent to the determination of the mean value of the 
sLDF.
As a first step, let us revisit a few well-known results in terms 
of the 
sLDF.

\subsection{Domain limits and average properties}\label{sub:lmax}

\begin{figure}
\resizebox{\hsize}{!}{\includegraphics{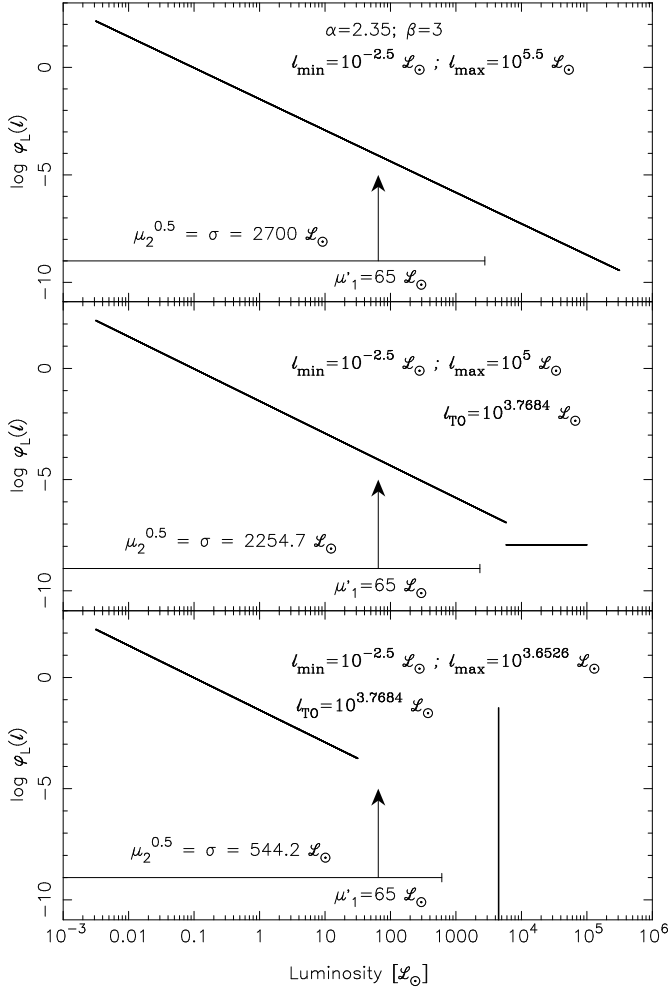}}
\caption[]{Schematic representation of the 
of the 
sLDF
for three different cases: main sequence only (top), main sequence plus a constant post main sequence (middle), and main sequence plus a bumpy post main sequence (bottom). The position of the mean ($\mu_1'$) and the region corresponding to $\mu_1' \pm 1 \, \sigma$ are explicitly  marked.}
\label{fig:LDFsimple}
\end{figure}

In synthesis modeling it is a well established fact that the integrated luminosity of the ensemble is dominated by the most massive stars in the cluster. In the following we will illustrate this point by means of three simplified scenarios that are representative of real sLDFs.
These are shown graphically in Fig.~\ref{fig:LDFsimple}. In each panel,
the position of the mean $\mu_1'$ and the region corresponding to $\mu_1' \pm 1 \, \sigma$ are explicitly  marked.

\paragraph{Case 1: Main sequence luminosity function}

In the first case let us assume that all the stars are in the main sequence (MS), and that 
$\ell \propto m^{\beta}$. Assuming a power-law IMF with index $-\alpha$, the 
sLDF
 can be expressed as:

\begin{equation}
\varphi_{\mathrm L} \propto \ell^{-\frac{\alpha}{\beta}} \cdot \frac{1}{\beta}\, \ell^{-\frac{\beta -1}{\beta}} = \frac{1}{\beta} \, \ell^{\frac{1-\alpha-\beta}{\beta}}.
\end{equation} 

\noindent The constant factor can be determined by imposing the normalization condition:

\begin{equation}
\varphi_{\mathrm L} =  \frac{1-\alpha}{\beta}\frac{1}{\ell_\mathrm{max}^{\frac{1-\alpha}{\beta}} - \ell_\mathrm{min}^{\frac{1-\alpha}{\beta}}}\,\ell^{\frac{1-\alpha-\beta}{\beta}} = \frac{A}{\beta} \,\ell^{\frac{1-\alpha-\beta}{\beta}}.
\end{equation} 

The mean value of the luminosity is then:

\begin{equation}
\mu_1'  = \frac{A}{\beta} \int_{\ell_\mathrm{min}}^{\ell_\mathrm{max}} \ell \cdot \ell^{\frac{1-\alpha-\beta}{\beta}} d\ell = 
\frac{A}{1+\beta-\alpha}\cdot \, \biggl(\ell_\mathrm{max}^{\frac{1+\beta-\alpha}{\beta}} - \ell_\mathrm{min}^{\frac{1+\beta-\alpha}{\beta}}\biggr).
\end{equation}

If $1+\beta-\alpha > 0$, the mean luminosity is driven by $\ell_{\mathrm{max}}$. In a typical situation with $\beta \approx 3$, the most luminous stars will dominate the luminosity if $\alpha < 4$: this is the case of Salpeter's IMF \citep{Sal55}.
If the IMF departs from a simple power law, the conclusions are less straightforward. For example, if we consider the log-normal IMF by \cite{MS79} and approximate it as a power-law series, a value of  $\alpha > 4$ is reached near 100 M$_\odot$ \citep[see Fig. 1 in][]{Kro01}. In this case, the dominant stars will not be the most massive ones.

\paragraph{Case 2: Luminosity function with a constant post main-sequence contribution}

Let us go a bit further and add a post-main sequence (PMS) contribution to the 
sLDF.
The 
sLDF
can now be divided in two regimes, corresponding to the MS and to the PMS respectively. As a first approximation, we will assume that the PMS phase gives a constant contribution to the 
sLDF,
implying that any point of the PMS portion of the isochrone is equally probable.
The 
sLDF
can be described by the expression:

\begin{equation}
\varphi_{\mathrm L} =  \Biggl\{
\begin{array}{ll}
    \frac{A}{\beta} \, \ell^{\frac{1-\alpha-\beta}{\beta}} ,   &  \text{if $\ell \, \in (\ell_\mathrm{min},\ell_\mathrm{TO})$},\\
    \frac{A}{1-\alpha}\, \frac{\ell_\mathrm{max}^{\frac{1-\alpha}{\beta}} - \ell_\mathrm{TO}^{\frac{1-\alpha}{\beta}} }{\ell_\mathrm{max} - \ell_\mathrm{TO} } &  \text{if $\ell \, \in (\ell_\mathrm{TO},\ell_\mathrm{max}$)},\\
\end{array}
\end{equation}

\noindent where $\ell_\mathrm{TO}$ is the turn-off luminosity. 
The absolute value of the PMS contribution is chosen so as to yield the same coefficient $A$ for the two regimes, while preserving the overall normalization;
that is, we are forcing the PMS to have a fixed weight with respect to the MS phase. This choice allows us to keep the complexity of the expressions to a minimum, but it has no influence on the general conclusions, which would be reached even if we dropped this assumption.

If we further assume that $\ell_\mathrm{TO}^{\frac{1+\beta-\alpha}{\beta}} >> \ell_\mathrm{min}^{\frac{1+\beta-\alpha}{\beta}}$, the mean value of the 
sLDF
is:

\begin{equation}
\mu_1'  \approxeq \frac{A}{1+\beta-\alpha} \ell_\mathrm{TO}^{\frac{1+\beta-\alpha}{\beta}} \, + \frac{A}{2(1-\alpha)}(\ell_\mathrm{max}^{\frac{1-\alpha}{\beta}} - \ell_\mathrm{TO}^{\frac{1-\alpha}{\beta}})(\ell_\mathrm{max} + \ell_\mathrm{TO}).
\end{equation}

\noindent That is, the mean depends on both $\ell_\mathrm{TO}$ and $\ell_\mathrm{max}$. Note that, since in the PMS phase the relation between the mass and the luminosity is not simple anymore, the above result cannot be easily expressed as a function of the initial mass. As a trivial example, if the age $t$ is larger than the lifetime of a star of initial mass $m^{\mathrm{up}}$, then $\ell(m^\mathrm{up},t) = 0$.

\paragraph{Case 3: sLDF
with a bumpy post main-sequence contribution}

As a final example, let us assume that the sLDF
has a narrow peak in addition to the MS contribution. This scenario describes the case in which PMS stars have all the same luminosity, as in the horizontal branch or in the red giant clump phase; this case can be modeled by adding a pulse function to the MS section of the 
sLDF. 
Since PMS luminosities are larger than MS luminosities, the pulse function is located at $\ell_\mathrm{max}$. The 
sLDF
is therefore:

\begin{equation}
\varphi_{\mathrm L} =  \Biggl\{
\begin{array}{ll}
    \frac{A}{\beta} \, \ell^{\frac{1-\alpha-\beta}{\beta}}    &  \text{if $\ell \, \in (\ell_\mathrm{min},\ell_\mathrm{TO})$,}\\
     \frac{A}{1-\alpha}\, (\ell_\mathrm{max}^{\frac{1-\alpha}{\beta}} - \ell_\mathrm{TO}^{\frac{1-\alpha}{\beta}})\, \delta(\ell - \ell_\mathrm{max})
      &  \text{if $\ell \, \in (\ell_\mathrm{TO},\ell_\mathrm{max}$)}.\\
\end{array}
\end{equation}

Again, we are forcing the PMS to have a fixed weight with respect to the MS phase for simplicity reasons. In this case, assuming again that  $\ell_\mathrm{TO}^{\frac{1+\beta-\alpha}{\beta}} >> \ell_\mathrm{min}^{\frac{1+\beta-\alpha}{\beta}}$:

\begin{equation}
\mu_1'  \approxeq \frac{A}{1+\beta-\alpha} \ell_\mathrm{TO}^{\frac{1+\beta-\alpha}{\beta}} \, + \frac{A}{1-\alpha}(\ell_\mathrm{max}^{\frac{1-\alpha}{\beta}} - \ell_\mathrm{TO}^{\frac{1-\alpha}{\beta}}) \, \ell_\mathrm{max}.
\end{equation}

\noindent So, again, the mean depends on both $\ell_\mathrm{TO}$ and $\ell_\mathrm{max}$. 

These three examples illustrate the general fact that the mean luminosity depends strongly on the value of $\ell_\mathrm{max}$. The dependence on $\ell_{\mathrm{max}}(t)$  would be even stronger for higher-order moments of the distribution. 
As \cite{GGS04} point out, the dependence of the results of synthesis models 
on the high-luminosity end of the 
sLDF
is so strong that, without such a limit, synthesis models could not even be computed!  
As a further example of the relevance of the upper limit of the 
sLDF,
we have shown in a previous paper that it also defines a 
lower limit for the luminosity of real clusters to be 
described by standard synthesis models \citep{CL04}.  

These examples also illustrate the following interesting facts:
{\it (i)} The mean value does not necessarily give 
information on the distribution of luminosities, to the extreme that there can be situations in which the probability to find a source in the region around the mean value is zero (Fig.~\ref{fig:LDFsimple}, bottom panel) . This is the opposite of a Gaussian distribution, in which the mean is also the most probable value.
{\it (ii)} Different distributions can have the same mean value but different variances. In fact, this circumstance permits to use surface brightness fluctuations (SBF: the ratio between the variance and the mean value of the luminosity function) to break the age-metallicity degeneracy, which makes clusters with different ages and metallicities have the same $\mu_1'$. 
{\it (iii)} The value of $\mu_1' - \sigma$ is negative in our three examples. This shows clearly that assuming, e.g., that $\mu_1' \pm 1 \sigma$ includes $\sim 68$\% of the elements of the distribution 
 can be grossly mistaken in the case of non-gaussian distributions.
As we will see later, this can also be the case with the distribution function of the integrated luminosity of an ensemble: this limits the use of goodness-of-fit methods based on the comparison with the mean, such as the $\chi^2$ test. 

Summing up, our schematic but realistic 
sLDFs
show that knowledge of the mean and the variance does not provide a handle on the problem if the shape of the distribution is not known. 
As \cite{GGS04} argue, the use of the mean value instead of the most probable one produces a systematic bias in the determination of integrated properties. 

\subsection{Implications of mass binning}\label{sub:binning}

Although Eq.~\ref{eq:muIMF} is expressed as an integral, synthesis codes
approach it through numerical approximations. There are several reasons for this:
first, the input data are available in tabular format rather than as 
analytical relations; e.g., given the enormous complexity of stellar evolution
it is not possible to derive stellar properties as analytical functions of, say, 
initial mass and stellar age. Rather, stellar tracks are computed for a discrete and finite 
set of mass values, and their properties are conveniently interpolated
{\it a posteriori}. 
Furthermore, calculated stellar tracks
are generally expressed in tabular form, although there have been sporadic attempts at expressing them in
analytical form \citep[e.g.][]{Touetal96}.
Second, even if analytical relations existed, their integration would
plausibly require numerical methods.
Third, the numerical precision of computers is limited.
These circumstances force the actual calculations of synthesis models 
to be numerical rather than analytical: as a consequence, the mass variable must be binned.
We will focus here on the implications of mass binning for Eq. \ref{eq:muIMF}.

Introducing binning, Eq. \ref{eq:muIMF} can be expressed as:

\begin{equation}
\mu_1'(t) = \int_{m^\mathrm{low}}^{m^{\mathrm{up}}} \ell(m;t) \, \varphi_\mathrm{M}(m) \, dm \simeq \sum_i  w_i \,\ell_i(t),
\label{eq:muwi}
\end{equation}

\noindent where the index $i$ identifies the mass bin and
$\ell_i(t)$ is the (time-dependent) luminosity of the $i$-th bin; the
approximation holds only if the luminosity $\ell_i(t)$ is
indeed representative of the whole mass bin.  The coefficient $w_i$
is computed by means of the expression:

\begin{equation}
w_i = \int_{m_i^{\mathrm{up}}}^{m_{i}^{\mathrm{low}}} \varphi_\mathrm{M}(m) \, dm,
\label{eq:wi}
\end{equation}

\noindent where $m_i^\mathrm{low}$ and $m_{i}^{\mathrm{up}}$ 
are the lower and upper limits of the $i$-th mass bin (the specific
way in which these limits are defined varies from code to code).  In
the framework of deterministic synthesis models, $w_i$ is
deterministically interpreted as the fraction of the total number of stars
enclosed in the given bin.  In the probabilistic approach, however,
such number is not fixed. Each star is either born within a given mass
bin $i$, with a probability $w_i$, or outside it. 
When $N_{\mathrm{tot}}$ stars are selected, the number of stars in each mass bin 
follows a binomial distribution. Furthermore, since all the bins share the same
number $N_{\mathrm{tot}}$ of stars, there is a finite covariance among bins:
that is, the collection of the mutually covariating binomial distributions
constitutes a multinomial distribution. According to this approach, $w_i$ also represents 
the mean (as opposed to {\it exact}) contribution of each bin to the total number of stars.
This interpretation is shared by statistical standard codes and Monte
Carlo codes.

\subsubsection{Distribution of stars in each bin}

The statistical strand of standard models was born with
the goal of devising statistical tools to be applied
to synthesis models. 
In this context, the binomial probability distribution 
of stars in individual bins had been approximated by a 
Poisson distribution to simplify its handling \citep{CVGLMH02}.
As is known, the Poisson approximation 
is accurate only when the number of events $N$ is large and
the probability $p$ is vanishingly small, in such a way that $Np$ stays finite:
we will show in the following that such approximation 
is not always accurate enough for our problem.

A first point to consider is that speaking of a distribution of stars in bins only makes sense if a value of $N_{\rm tot}$ is considered. In both cases considered, Poisson and binomial, the mean value of the number of stars in a bin is $\mu_i(n_i) = N_{\rm tot} \times w_i$.
Let us now illustrate the difference between the two distributions for the case of $N_\mathrm{tot} = 1$. The probability for the source to fall in the i-th mass bin is,
in the Poisson approximation,  $p_i^{P}(1)=w_i \, e^{-w_i}$. In the binomial case such probability is
$p_i^{b}(1)=w_i$, which is the correct value. The difference is non-negligible for large
$w_i$ values, which is typical for bins at small masses.

Furthermore, when $N_{\rm tot}$ stars are considered,
the Poisson approximation predicts the ratio $M_2/M_1'$ of the variance to the 
mean to be unity,
whereas in Monte Carlo simulations this ratio shows a trend 
toward smaller values for small bin masses  \citep[Fig. 1 of][]{CVGLMH02},
revealing the underlying binomial distributions, in which:

\begin{equation}
 \frac{M_2}{M_1'} =  \frac{N_{\rm tot}  w_i(1-w_i)}{N_{\rm tot}  w_i }= 1 - w_i.
\end{equation}

\noindent  Note that, as expected, the discrepancy with respect to 1 grows
  larger for larger $w_i$ values  -- i.e., bins at lower masses.
  Unfortunately, \cite{CVGLMH02} failed to interpret
  this result in term of a binomial distribution,
  and proposed to solve the discrepancy by reducing 
  the size of the bins, while keeping the Poisson approximation.
  Of course, reducing the size of the bin (or, equivalently, $w_i$) 
  leads to a closer similarity between the two distributions,
  but this solution is not always viable due to the limited
  resolution of the input ingredients. A further point
  is that, because $N_\mathrm{tot}$ is shared by all the bins, 
  the numbers of stars in different bins covariate:
  while this effect is neglected when the bin distributions are modeled
  by independent Poisson distributions, it arises naturally 
  when a multinomial distribution is used.

The difference between the Poissonian approximation and the multinomial description is particularly clear if we compute the variance of the 
pLDF.
Consider the expression for the variance obtained assuming that 
the number of stars in each bin 
is described by a Poisson distribution, 
and that there is no correlation between bins
\citep[e.g.][among others]{CVGLMH02}:

\begin{eqnarray}
M_2 &=& \sum_i  \mu_2(n_i) \,\ell_i^2 =   N_{\rm tot}  \, \times \,  \sum_i  w_i \,\ell_i^2.
\label{eq:varwiP}
\end{eqnarray}

\noindent Now, consider the expression corresponding to a multinomial distribution, where, by definition, $\mathrm{cov}(n_i,n_j) = - N_\mathrm{tot} \, w_i\, w_j$:

\begin{eqnarray}
M_2 &=&   \sum_i \mu_2(n_i) \ell_i^2  +  \sum_i \sum_{j \neq i} \ell_i \ell_j \, \mathrm{cov}(n_i,n_j) = \nonumber \\
     &=&  N_{\rm tot}  \, \times \, \biggr( \sum_i w_i \ell_i^2 - \sum_i (w_i \ell_i)^2  \biggl) - \nonumber\\
     &  & - N_{\rm tot}  \, \times \, \biggr( \sum_i \sum_{j\neq i} \ell_i \ell_j \, w_i w_j \biggl).
\label{eq:varwiM}
\end{eqnarray}

Comparing Eq.~\ref{eq:varwiP} to Eq.~\ref{eq:varwiM}, it can be seen that
the latter contains two additional terms:
the term $N_{\rm tot}  \, \times \, \sum_i (w_i \ell_i)^2$ arises as a result of
dropping the Poisson approximation, 
and the term $\sum_i \sum_{j \neq i} \ell_i \ell_j \, \mathrm{cov}(n_i,n_j)$
represents the mutual dependence of bins, and hence arises {\it as a direct consequence of binning}. 

These considerations show that the Poisson approximation, motivated by simplicity of handling, may break down in our problem \citep[see also][]{L00}.
While the distinction between the Poisson approximation and the multinomial description is important in order to make sense of Monte Carlo simulations, 
it is fundamental in view of the discussion
on our proposed 
probabilistic
formalism (Sect.~\ref{sect:probabilistic}).

\begin{figure}
\resizebox{\hsize}{!}{\includegraphics{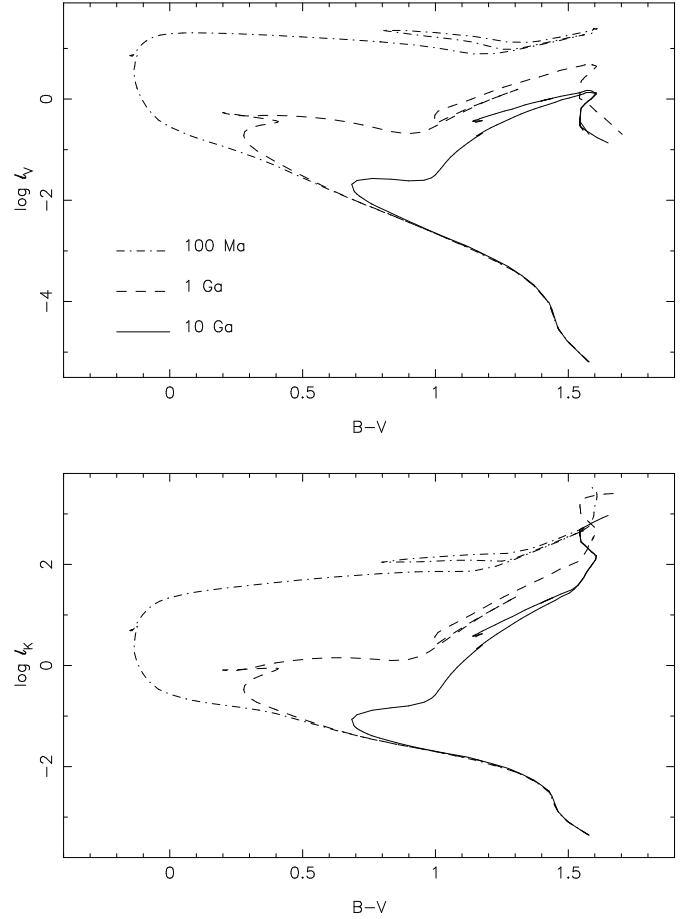}}
\caption[]{Top: isochrones in the $\ell_{\mathrm V}$ vs. B-V color-magnitude diagram for three different age values computed by \cite{Gi02} based on the evolutionary tracks by \cite{MG01}. Bottom: same as above, for $\ell_{\mathrm K}$.}
\label{fig:HR}
\end{figure}

\subsection{Fast evolutionary phases}\label{sub:fast}

\begin{figure*}
\resizebox{\hsize}{!}{\includegraphics{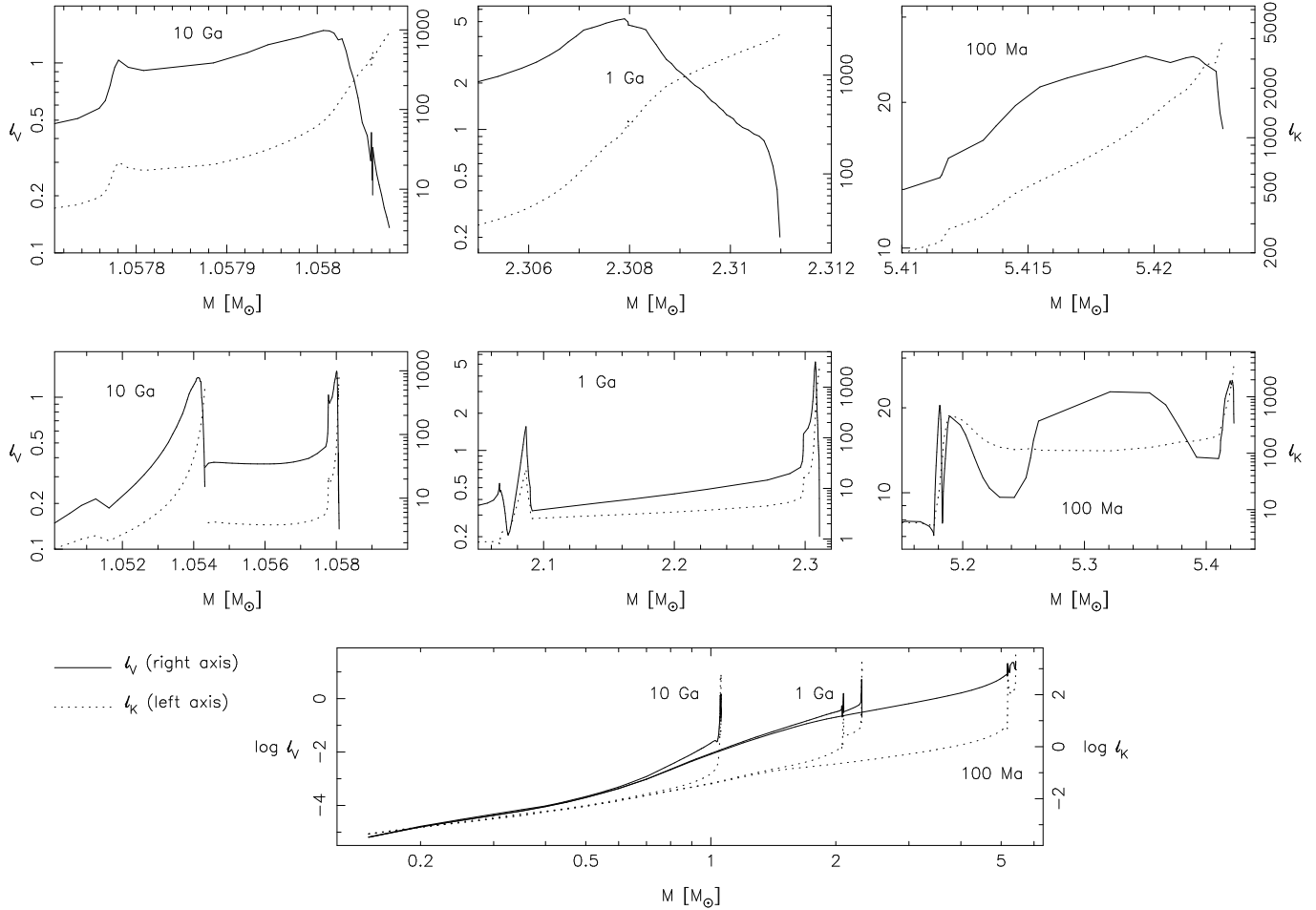}}
\caption[]{Details of fast evolutionary phases in the V (solid line, left axis) and K (dotted-line, right axis) bands. Bottom panel: complete isochrones. Middle panels: blow-up of the mass axis, showing details of fast evolutionary phases at three different ages. Top panels: same as middle panels, with a more extreme blow up. The isochrone set is the same of  Fig. \ref{fig:HR}.}
\label{fig:fast}
\end{figure*}

In writing Eq.~\ref{eq:ldf-imf}, we have tacitly assumed that the luminosity 
is a well-behaved function of the mass so that the isochrone is always defined.
However, this condition is in fact often violated, since
a typical isochrone features shallow sections as well as peaks and discontinuities
in the $(\ell, m)$ plane.
Shallow sections correspond to quiescent phases of stellar evolution, 
where evolution is slow 
(e.g. the MS); peaks correspond to faster phases 
(e.g. the asymptotic giant branch, AGB);
and discontinuities correspond to abrupt transitions between phases
(e.g. the onset of central helium burning in intermediate mass stars)
or jumps in stellar behavior (the transition between WR and non-WR-type structures).
In the following, peaks and discontinuities will generically be referred to
as {\it fast evolutionary phases}. 
Some of these cases are illustrated by the comparison of Fig.~\ref{fig:HR}
with  Fig. \ref{fig:fast}.
In Fig. \ref{fig:HR} the isochrones computed by \cite{Gi02}
based on the  evolutionary tracks by \cite{MG01}\footnote{These models are available at {\tt http://pleiadi.pd.astro.it/}.}
are shown, at selected age values.
Fig. \ref{fig:fast} shows the relation between the initial mass 
and the luminosity in the V and K bands for the same isochrones,
focusing on fast evolutionary phases.

If the mean value of the 
sLDF
 is determined through 
the numerical approximation of Eq.~\ref{eq:muwi},
fast evolutionary phases are difficult to handle,
because a small difference in initial stellar mass yields a large
difference in luminosity, so that the result of the numerical integration
crucially depends on which luminosity is chosen to be
representative of a given mass bin.
In principle, this difficulty can be dealt with
by decreasing the width of the mass bin and choosing mass
bins that do not go across discontinuities; however,  
the available resolution in mass is governed by the evolutionary tracks
used by the code, and
is generally too low to resolve adequately such phases in synthesis models.

This is a severe problem, since fast evolutionary phases are ubiquitous in 
post main-sequence evolution, and at certain frequencies 
they bear a major weight in the luminosity balance. 
Unfortunately, the way in which this problem is
tackled is often labeled a `technical
detail' of the computation and dismissed as unimportant,
and thus the papers describing evolutionary synthesis models
do not generally make any reference to its solution -- in spite of
its difficulty and of the potentially disastrous consequences of
incorrect assumptions.  Here are a few examples of the ways in which this
problem has been approached: a) in the Starburst99 synthesis
code \cite[][]{SB99}, for certain metallicity values,  
an undocumented stellar track at 1.701 M$_\odot$ 
is added to the tabulated track of 1.70 M$_\odot$ by
\cite{Schetal92} and \cite{Schetal93}, to avoid the mass bins
going across the discontinuity of the stellar models' behavior
at such mass (C. Leitherer, D. Schaerer, \& G.
Meynet, private communication); b) to deal with the same problem, 
additional evolutionary tracks around the same mass range 
are used in the computation of the isochrones
by \cite{Casetal03} and \cite{Caretal04} (S. Degl'Innocenti, private communication)
and \cite{Broetal99} (E. Brocato, private communication); 
c) to avoid the intrinsic discontinuity in the isochrones, 
the same mass is used twice in the isochrones by \cite{Gi02}, 
namely at the end of the red giant branch and 
at the beginning of the horizontal branch 
(S. Bressan, private communication).

As a final point, consider that fast evolutionary phases 
are those stages in which the evolution
is rapid in any of the relevant luminosity bands,
and not only the bolometric luminosity.
Therefore, an isochrone that is adequately sampled
in one band is not necessarily so in all the bands.

\subsection{Transient phases and fuzzy stellar behavior}\label{sub:trans}

Traditional synthesis models rely on the possibility to express the luminosity
as a function of mass (Eq.~\ref{eq:ldf-imf}), so that the integrals over luminosity are
transformed into integrals over mass. However, there are 
stellar phases in which this is impossible to do, because the luminosity at a given
age is not univocally determined by the mass. A few examples are:
the thermal pulse phase in AGB stars, which is poorly resolved by stellar evolutionary models;
supernova (SN) light curves, in which a star of fixed mass spans a large range of luminosity
almost instantaneously; variable stars; rotating stars, in which the luminosity 
emitted depends on both the rotation velocity and the inclination angle.
Partly
because of this difficulty, these 
phenomena 
are generally not included in synthesis models;
whereas approaching the problem 
from the point of view of distribution functions
permits to account for them by including them in the sLDF - provided they can be modeled quantitatively.
This point will be developed further in Sect.~\ref{sub:uncertainties}.
See also  \cite{GGS04} for an example 
on the inclusion of variability.

Of course, in many cases of interest the modeling of the phenomenon 
may be a difficulty in itself.
For example, modeling rotation is in itself problematic.
Rotating stellar models are just beginning to appear on the market, and the distribution of velocities in a stellar population is largely unknown. Stellar rotation is not yet included in synthesis models primarily because of these uncertainties, and not only because synthesis models are not flexible enough.
But if stellar rotation or any other distributed stellar behavior is properly understood, or at least satisfactorily modeled, our method will permit to include it in synthesis modeling.

\subsection{Implementation of model atmospheres}\label{sub:atmosphere}

Usually, the available model atmospheres form a coarse grid in the
(log $g$, log $T_{\rm eff}$) plane, whereas isochrones are
generally continuous in the plane. In order to assign 
a model atmosphere to each isochrone location, 
one can either choose the nearest atmosphere of the grid,
or interpolate between nearby atmospheres. Assigning the nearest
atmosphere implies a further binning of data and may originate
jumps in the results, although it is often assumed that these jumps 
cancel on average when a whole population is considered.
This problem will not be further discussed here; a more extensive discussion
can be found in \cite{CL05}.

In the next section, we will review the ways in which the basic task of
synthesis codes is accomplished, given the difficulties outlined here.

\section{Computational algorithms}\label{sect:algorithms}

This section outlines the basic ways in which the task
of computing the luminosity is specifically performed by standard codes.
The first two parts, \ref{sect:iso_synthesis} and \ref{sect:FCT},
describe techniques implemented in standard models. The third part,
\ref{sect:MC}, describes how the Monte Carlo method is put into practice.

\subsection{Isochrone synthesis}\label{sect:iso_synthesis}

The commonest technique to compute the 
luminosity of stellar populations
is known as isochrone synthesis. Isochrone synthesis
is the numerical integration of the product
between the IMF and the isochrone 
approximated as in the rightmost part of Eq.~\ref{eq:muwi}:
$\mu_1'(t) = \sum_i  w_i \,\ell_i(t)$.
The strongest assumption of this numerical approximation is 
that each $\ell_i(t)$ is representative
of all the stages included in it. 
Fast evolutionary phases evidently 
pose a problem in the integration, since a small
interval in mass can correspond to very different luminosity values.
In isochrone synthesis, the problem is solved 
increasing the number of mass bins that map fast evolutionary phases. 
However, 
this strategy 
eventually requires assuming mass bins narrower 
than the precision of the published tracks and isochrones. 
It is interesting to note that one of the
main advantages of isochrone synthesis mentioned by \cite{CB91} is
that it produces `smooth' results; however, this advantage
only hides the problem of fast evolutionary phases, but does not
solve it. 

\subsection{The fuel consumption theorem}\label{sect:FCT}

As an alternative method, one can avoid using the expression
of Eq.~\ref{eq:muwi}, and 
perform instead
the integration in the luminosity domain, that is 
integrate $\ell \, \varphi_\mathrm{L}(\ell)$: 
in this way, 
the fast evolution
of luminosity is automatically taken care of. 
This is the basis of the so-called fuel
consumption theorem (FCT):
to understand fully the FCT and its assumptions,
we refer the reader to the original paper \citep[][see also \citeauthor{Mar98} \citeyear{Mar98}]{RB86,Buzz89}.

Note that isochrone synthesis and the FCT can be coupled: the computation of isochrones can be done so as to fulfill the FCT requirements. We refer to \cite{MG01} for an extended discussion on the subject \citep[see also][]{Breetal94}.

\subsection{Monte Carlo methods}\label{sect:MC}

The Monte Carlo method can be implemented in different ways, 
which take advantage of its potential to varying degrees, 
so they are not completely equivalent. To understand this point,
consider how a sample of stars selected through a Monte Carlo
mass assignment can be handled:
first, the stars can be either followed individually throughout
their evolution, or grouped in mass bins characterized by average properties;
second, in either case an atmosphere, used to transform the bolometric luminosity into 
observable quantities,  can be assigned by either choosing 
the nearest model in the available grid,
or interpolating between nearby atmospheres (see~\ref{sub:atmosphere}). 
Therefore there is a total of four ways in which 
a Monte Carlo-selected stellar population\footnote{A
  further degree of freedom in Monte Carlo simulations is that they can be performed
  by either fixing the number of stars or the total stellar mass. This aspect will not be discussed here, and we will consider only Monte Carlo simulations with a fixed number of stars.}
can be treated.
In the literature, there are examples of at least three  of these:

\begin{itemize}
\item[{(i)}] The stars obtained through the Monte Carlo selection
  are grouped in bins; to each bin, the nearest available
  atmosphere model is assigned, which depends on the age considered.
  That is, the total luminosity is computed by means of the expression

\begin{equation}
{\cal L} =  \sum_i  n_i \,\ell_i(t),
\label{eq:muwi_MC}
\end{equation}

 \noindent where $n_i$ is the number of stars that 
  have fallen into the i$-th$ mass bin, and $\ell_i$ 
  is the luminosity of the atmosphere assigned to the bin;
  Eq.~\ref{eq:muwi_MC} is analogous to Eq.~\ref{eq:muwi}, with the only difference that
  the $n_i$ values vary from simulation to simulation.
  This strategy is followed by \citet{Brutuc,BC03} (G. Bruzual, private
  communication). This implementation of the MC method is not very time-consuming, 
  so it
  is quite effective in obtaining an overview of the effect of sampling 
  in different observables, which is in fact the goal of the quoted papers. 
  It has, however, two important disadvantages: the mass binning hinders a proper
  sampling of the discontinuities in the isochrones, and the $\ell_i$ assignation
  may either smooth out or spuriously amplify transient spectral features 
  (see Sect.~\ref{sub:atmosphere}). 

\item[{(ii)}] The stars obtained through the Monte Carlo procedure
  are followed individually throughout their evolution, but the
  luminosities are assigned as in the previous method, 
  that is choosing the nearest model in the grid. 
  This approach, although similar to isochrone synthesis, is
  fundamentally different in that it avoids the additional
  rebinning implicit in the isochrone synthesis method. Models of
  this kind have been presented, for example, in \citet{MHK91,CMH94,KFAF99,Cgam00}.
  These models work better than the previous at mapping discontinuities
  in the isochrones \cite[i.e., the value of $d\ell/dm$ is better evaluated:
  see, e.g.  Fig. 4 of][]{MHK91}, but present the same disadvantages
  with respect to the assignment of $\ell_i$.
  
\item[{(iii)}] A further possibility is to perform Monte Carlo simulations 
  over the luminosity function: the Monte Carlo-sampled stars are 
  followed individually, and each is assigned a tailored atmosphere.
  Doing so requires performing interpolations in the $\ell_i$ grid,
  and permits reproducing the path in the HR diagram of individual stars.
  The individual luminosity values obtained in this way are eventually
  summed to yield the total luminosity.
  These models exploit the potential of Monte Carlo method to its 
  maximum and are the only ones able to map correctly the luminosity function
  without the necessity of binning. A sufficiently numerous set of 
  Monte Carlo simulations of this kind provides directly the 
  distribution function of the ensemble luminosity. The bleeding edge 
  of this method lies, of course, in the interpolations techniques 
  used both in the stellar evolution and in the atmosphere assignment.
  Examples of applications of this method can be found in \citet{CRBC03,CVG03,Gproc00},
  among others. 
  
\end{itemize}

\section{Old tools for a new approach: the probabilistic formulation}\label{sect:probabilistic}

The conclusions arising from this  brief overview
of population synthesis can be summarized in the
following points:

\begin{itemize}

\item Deterministic standard models are based on a misunderstanding
  of the computed quantities. While it is generally claimed that they compute 
  the integrated luminosity of a model cluster, they in fact
  compute the mean luminosity of the 
sLDF.

\item Statistical standard models (i.e., those that compute the 
mean and the variance of the luminosity distribution)
give the correct interpretation, but they
have a limited interpretative power. Knowledge of 
the mean and variance is not enough to characterize a distribution, much less to 
explain its shape in physical terms.

\item Suitably-done Monte Carlo simulations have the potential to 
  bypass most of the problems of population synthesis arising
  as a consequence of the statistical nature of the problem.  
  However, they are extremely expensive in terms of CPU time and 
  disk storage space, and require a considerable human effort to
  be analyzed. Furthermore and most importantly, the analysis of a set of Monte Carlo 
  simulations performed without a grasp of the underlying statistics would be purely phenomenological,
  precluding the possibility to generalize the results to further simulations or real clusters. 
   
\end{itemize}

In this section, we will describe a 
probabilistic
formalism
that automatically frames the problem in its most natural interpretation,
and opens the way to a deep understanding of the underlying physical problem.
The first subsection is devoted to recall standard statistical 
concepts for those readers without a background in statistics
and probability theory. 
Those who are already expert in this field can skip this part.

A note on definitions is due here. Throughout this paper, we adopt the conventional (though often overlooked) distinction between probability theory and statistics. A {\it statistical formulation} is one that seeks to intepret a sample of experimental data in terms of an underlying distribution. A {\it probabilistic} formulation is one that assumes an underlying distribution and uses it to predict the resulting distribution of experimental data. The traditional Monte Carlo method has followed a statistical approach: conclusions were drawn from the observation of a spread in the results of simulations. The present paper lays the foundations for a probabilistic approach, in that it seeks to give a formal description of the underlying distributions, and to make quantitative predictions based on them.
Obviously, the two approaches complement each other and partially overlap; concepts like the one of distribution function belong to both probability theory and statistics, thus we will refer to them as probabilistic or statistical alike. When it comes to define our method with respect to existing ones, however, we will systematically draw a distinction and refer to it as a probabilistic formalism. 

\subsection{Basic statistical concepts}

As we have seen in Sect.~\ref{sect:overview}, the luminosity
of stars can be characterized in terms of an underlying
sLDF
(Eq.~\ref{eq:Ltot_mean}). The exploitation 
of this approach permits finding an effective
solution to the problem of computing the integrated 
properties of stellar populations.
The formalism presented here to this scope
is not new in terms of 
theory of distributions
and can
be found in any advanced textbook of statistics \cite[e.g.][]{KS77}.
For completeness, we briefly review here the relevant concepts.

The properties of the 
sLDF 
as a distribution can be studied by 
computing its moments; the first moment is the mean,
which we already encountered in Eq.~\ref{eq:Ltot_mean}:

\begin{equation}
\mu_1' = \int_0^\infty \ell \, \varphi_\mathrm{L}(\ell) d\ell.
\label{eq:Lmean}
\end{equation}

\noindent 
The general definition of the n-th moment of the 
sLDF
is the following:

\begin{equation}
\mu_n(a) = \int_0^\infty (\ell-a)^n \, \varphi_{\mathrm{L}}(\ell) \, d\ell.
\end{equation}

\noindent If $a=0$, we call it `raw moment', while if $a=\mu_1'$ 
we call it `central moment'; in particular, 
the mean luminosity, which is the main output of synthesis models,
is the raw moment of 1st order.

In the following, we will adopt the notation $\mu_n$ to indicate central moments
and the notation $\mu'_n$ to indicate raw moments.
Let us write down explicitly the expressions for the second central moment (or variance) and the second raw moment of the 
sLDF,
and their mutual relation:

\begin{eqnarray}
  \mu_2'& =  & \int_0^\infty \ell^2 \, \varphi_{\mathrm{L}}(\ell) \, d\ell, \label{eq:2nd_momentp}\\
  \mu_2 & =  & \int_0^\infty (\ell - \mu_1')^2 \,  \varphi_{\mathrm{L}}(\ell) \, d\ell = \nonumber\\
        & =  & \mu_2' - \mu_1'^2;
 \label{eq:2nd_moment}
 \end{eqnarray}

\noindent
and, analogously, for the third and fourth moments:

\begin{eqnarray}
  \mu_3'& =  & \int_0^\infty \ell^3 \,  \varphi_{\mathrm{L}}(\ell) \, d\ell, \\
  \mu_3 & =  & \int_0^\infty (\ell - \mu_1')^3 \,  \varphi_{\mathrm{L}}(\ell) \, d\ell = \nonumber \\
        & =  & \mu_3' - 3 \mu_1' \mu_2' + 2 \mu_1'^3,\\
  \mu_4'& =  & \int_0^\infty \ell^4 \,  \varphi_{\mathrm{L}}(\ell) \, d\ell, \\
  \mu_4 & =  & \int_0^\infty (\ell - \mu_1')^4  \,  \varphi_{\mathrm{L}}(\ell) \, d\ell = \nonumber\\
        & =  & \mu_4' - 4 \mu_1' \mu_3' + 6 \mu_1'^2 \mu_2'  - 3 \mu_1'^4.
 \label{eq:3-4th_moment}
  \end{eqnarray}

Let us also introduce the characteristic function of the 
sLDF,
$\phi(p)$, that is its Fourier transform:

\begin{equation}
\phi_\mathrm{L}(p) = \int_{0}^\infty e^{ip\ell} \,  \varphi_{\mathrm{L}}(\ell) d\ell.
\label{eq:charfsLDF}
\end{equation}

The characteristic function has the following properties: 
the coefficients of its Taylor expansion in $ip$ are the raw moments of the
distribution, while the
coefficients of the Taylor expansion in $ip$ 
of its logarithm $\ln \phi_\mathrm{L}(p)$ are the so-called 
{\it cumulants} of the distribution, $\kappa_n$:

\begin{equation}
\ln \phi_\mathrm{L}(p) = \sum_{r=0} \kappa_r \, \frac{(ip)^r}{r!}.
\end{equation}

\noindent It follows that moments and cumulants are related by the expression:

\begin{equation}
\sum_{n=0} \mu_n' \, \frac{(ip)^n}{n!}  =  \exp \, \biggl({\sum_{r=0} \kappa_r \, \frac{(ip)^r}{r!}}\biggr).
\label{eq:kappadef}
\end{equation}

\noindent In particular, the relations between the first four moments and cumulants are the following:

\begin{eqnarray}
\kappa_1 & = & \mu_1',   \label{eq:kappa1}\\
\kappa_2 & = & \mu_2   = \mu_2' - \mu_1'^2, \\
\kappa_3 & = & \mu_3   =  \mu_3' - 3 \mu_1' \mu_2' + 2 \mu_1'^3,\\
\kappa_4 & = & \mu_4 - 3 \mu_2^2  =  \mu_4' - 4 \mu_1' \mu_3' -3 \mu_2'^2 +12 \mu_1'^2 \mu_2'  - 6 \mu_1'^4.
\label{eq:kappa4}
\end{eqnarray}

An important property of cumulants is that they are independent of the
assumed origin of the distribution, except for $\kappa_1$:
they are also called sometimes ``semi-invariants'' due to this
property. If the origin is taken at the mean of the distribution,
$\kappa_1 = 0$.

Finally, the skewness, $\gamma_1$,  and the kurtosis, $\gamma_2$, of the distribution 
are defined through ratios of the third and the fourth central moments respectively
to appropriate powers of the variance:

\begin{eqnarray}
\gamma_1 &= & \frac{\mu_3}{\mu_2^{3/2}} = \frac{\kappa_3}{\kappa_2^{3/2}}, \label{eq:gamma1} \\
\gamma_2 &= & \frac{\mu_4}{\mu_2^{2}} - 3= \frac{\kappa_4}{\kappa_2^{2}} . \label{eq:gamma2} 
\end{eqnarray}

\noindent (Note that the definition of skewness and kurtosis
may vary from author to author; alternative definitions can be found at
{\tt http://mathworld.wolfram.com}.) 
These two quantities enclose information on the shape of the
distribution: the skewness gives an idea of
how asymmetric the distribution is, and it can be related to 
the difference between the value of the mean and the mode 
(the most probable value).
The kurtosis is a measure of peakedness, i.e. of the symmetric deformation of the
distribution with respect to a Gaussian.
In particular, Gaussian distributions have $\gamma_1=0$
and $\gamma_2 = 0$; flatter distributions have negative kurtosis values,
while peaked distributions have positive kurtosis values.

\subsection{From stellar to population luminosity functions}\label{sect:s2pLDF}

In the previous section we have characterized the properties
of the luminosity function of individual stars.
Let's see now how the luminosity function of an ensemble of $N_{\mathrm{tot}}$
sources can be computed.

\subsubsection{Obtaining the pLDF: exact solution}\label{sub:exact_pLDF}

As a general rule, the PDF of an ensemble of 
variables is obtained as the convolution of the PDFs of
the individual variables. For example, let $\varphi_\mathrm{x}(x)$ be the PDF of a variable $x$ and $\varphi_\mathrm{y}(y)$ the PDF of a variable $y$ independent of $x$. The probability density of a variable $u=x+y$ is given by the product of the probabilities of $\varphi_\mathrm{x}(x)$ and $\varphi_\mathrm{y}(y)$ summed over all the combinations of $x$ and $y$ such that $u=x+y$:

\begin{equation}
\varphi_\mathrm{u}(u) = \int_{-\infty}^\infty \varphi_\mathrm{x}(z) \, \varphi_\mathrm{y}(u-z) \, dz = \varphi_\mathrm{x}(x) \,\otimes \,\varphi_\mathrm{y}(y),
\end{equation}

\noindent which is the definition of convolution. In our case, we are assuming that all the stars have luminosities distributed  following the same distribution function, $\varphi_\mathrm{L}(\ell)$,
and that the stars are independent on each other. Therefore, the PDF of an ensemble of $N_{\mathrm{tot}}$ stars is obtained
by convolving $\varphi_\mathrm{L}(\ell)$ with itself $N_{\mathrm{tot}}$ times:

\begin{equation}
\varphi_{\mathrm{L_{tot}}}({\cal L}) = \overbrace{ \varphi_{\mathrm{L}}(\ell) \otimes \varphi_{\mathrm{L}}(\ell) \otimes\, ... \, \otimes \varphi_{\mathrm{L}}(\ell)}^{N_\mathrm{tot}}.
\end{equation}

Hence, if the sLDF is known, the pLDF of an ensemble of $N_\mathrm{tot}$ stars can be computed by means of a convolution.
The convolution is conceptually straightforward, but it poses severe numerical problems.
The reason is the following. The convolution must be performed linearly in luminosity and,  in the general case, the dynamic range in luminosities spans eight orders of magnitude, from $10^{-2}$ L$_\odot$ to $10^{6}$ L$_\odot$. On the other hand, most programs that perform convolutions are based on Fourier transform routines that require a set of points ordered regularly on the $x$-axis; each time a convolution is performed the number of points on the $x$-axis is doubled. So, for a resolution of, say, 
0.01 L$_\odot$ (necessary to resolve
the low end of the luminosity function) $10^{8}$ points would be needed to define the luminosity function, and this number would be doubled each time a convolution is performed. Therefore, the points necessary to compute even a very undersampled population (a moderate number of convolutions) diverge rapidly, making it numerically unfeasible. 
We do not know any computational routine powerful enough to perform this task, and any feedback from the community about this subject is highly welcome.

An alternative solution is making the convolution logarithmically in the Fourier space (see next section). However, the numerical computation of the Fourier transform of a function with an irregular sampling is also difficult and, again, we haven't find any routine to perform it satisfactorily.

\subsubsection{Scaling properties of the LDF}\label{sub:scaling}

Convolutions in the normal space are equivalent to products in the Fourier space, so that:

 \begin{eqnarray}
\phi_{\mathrm{L_{tot}}}(P) &=& \phi_{\mathrm{L}}(p)^{N_\mathrm{tot}}, \label{eq:char_pop}\\
\ln \phi_{\mathrm{L_{tot}}}(P) &=& N_\mathrm{tot} \times \ln \phi_{\mathrm{L}}(p) =  N_\mathrm{tot} \times \sum_{r=0} \kappa_r \, \frac{(ip)^r}{r!}.\label{eq:ln_char}
\end{eqnarray}
 
Hence, 
the cumulants of the luminosity distribution 
of an ensemble of $N_{\mathrm{tot}}$ sources, $K_n$,
can be easily obtained 
from the cumulants of the 
sLDF,
$\kappa_n$, through the simple scale relation:

\begin{equation}
K_n = N_{\mathrm{tot}} \,  \times \, \kappa_n.
\label{eq:scale}
\end{equation}

\noindent Since $\kappa_1=\mu_1'$, this also implies that:

\begin{equation}
M_1' = N_{\mathrm{tot}} \,  \times \, \mu_1',
\label{eq:mu_scale}
\end{equation}

\noindent where $M_1'$ is the mean value of the distribution that describes the
luminosity of an ensemble of $N_{\mathrm{tot}}$ sources:
in other words, the mean luminosity 
obtained by synthesis models is scalable to a cluster of any size
-- including one with only 1 source!
 
Thus, 
probabilistic
reasoning confirms the intuitive expectation that
the properties of an ensemble of $N_{\mathrm{tot}}$ stars can be
obtained by direct scaling of the properties of the 
sLDF.
This result is fundamental for the interpretation of the output of
synthesis models in terms of real clusters. However, it is clear from
the above derivation that {\it this simple scaling rule only applies to cumulants,
  not to moments}; it is cumulants that hold the scale relations
  between the properties of the 
sLDF
and the distribution of the total
  luminosity of an ensemble. But, by virtue of Eqs.~\ref{eq:kappa1}$-$\ref{eq:kappa4}, cumulants
can be related to moments: $\kappa_1$ is equal to
the first raw moment, and the following 
$\kappa$s are equal to the central moments of the same order,
which can in turn be expressed in terms of the raw moments: 
therefore, 
to characterize a distribution
we can refer either to the moments
or to the cumulants. 

It is immediately seen from Eqs.~\ref{eq:gamma1} and~\ref{eq:gamma2}
that the shape of the distribution of the ensemble, when expressed in normal form (through a transformation of the distribution function to one with zero mean and unit variance: ${\cal L} \rightarrow x = {({\cal L}-M_1')}/{\sqrt{M_2}}$),  can be easily 
related to the shape of the 
sLDF:

\begin{equation}
\Gamma_1  =  \frac{1}{\sqrt{N_{\mathrm{tot}}}}\, \gamma_1, \label{eq:g1_G1}\\
\end{equation}
and 
\begin{equation}
\Gamma_2  =  \frac{1}{N_{\mathrm{tot}}} \,\gamma_2, \label{eq:g2_G2}\\
\end{equation}

\noindent where $\Gamma_1$ and $\Gamma_2$ are the skewness and the kurtosis
of the distribution of the ensemble.
Note that, in agreement with the central limit theorem, $\Gamma_1
\rightarrow 0$ and $\Gamma_2 \rightarrow 0$ for 
large enough $N_{\mathrm{tot}}$ values, i.e.  the distribution tends to a Gaussian.

Although the previous relations are useful to unveil the scale properties of LDFs,
knowledge of the moments of a distribution is useful but not sufficient 
to analyze it if its shape is unknown. For most application, one needs to know
whether the distribution can be approximated by a Gaussian, and in case it is not,
which its shape is. The following section will deal with the problem of characterizing a distribution by means of its cumulants. The technique suggested can be used
to solve two different kind of problems:
on one hand, it can be used to generate a theoretical pLDF from a sLDF when the convolution is not feasible. On the other, it can be used to infer the pLDF of an observed population.

\subsubsection{Obtaining the pLDF: approximate solution}\label{sub:appr_pLDF}

Alternative solutions go through obtaining an approximate expression for the pLDF.
A quantitative characterization of the pLDF by means of its cumulants can be obtained by means of approximate expressions.
To this aim, we suggest using the Edgeworth's series, which can be written schematically as:

\begin{eqnarray}
\varphi(x) = Z(x) \, \Biggl[ 1 + \sum_{i=1}^\infty t_i \Biggr],
\end{eqnarray}

\noindent
where $x$ is the normalized luminosity defined above, $Z(x)$ is the Gaussian distribution function, and the terms $t_i$ are obtained by the Chebyshev-Hermite polynomials multiplied by powers of the cumulants \citep{BM98}. This series is a true asymptotic series, i.e. the error is controlled when the series is truncated to a finite number of terms $n$. As \cite{BM98} demonstrate, the error is on the same order of the last term of the sum, $t_n$. When the error is small, i.e. the approximation is satisfactory, the term $\Sigma_n \equiv \sum_{i=1}^n t_i$ measures the deviation of the LDF from gaussianity.

\begin{figure}
\resizebox{8.5cm}{!}{\includegraphics{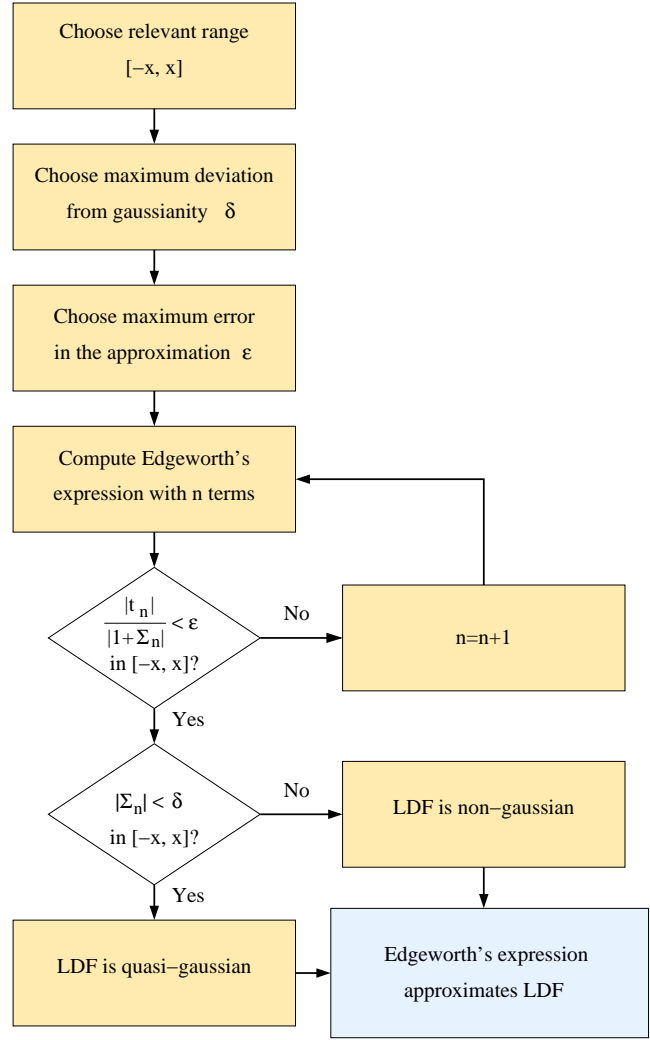}}
\caption[]{Algorithm to obtain an approximate expression for a pLDF based on the Edgeworth's series.}
\label{fig:edgeworth1}
\end{figure}

These properties can be used to obtain an explicit description of the distribution and to estimate its degree of gaussianity. The algorithm to be followed is described here and represented in Fig.~\ref{fig:edgeworth1}:

\begin{itemize}

 \item[{(i)}]The range of interest in x (i.e. the normalized luminosity) must be defined. This is necessary because convergence is first reached at small x, and propagates outward as more terms are included in the truncated series.
 \item[{(ii)}]The maximum deviation $\delta$ from gaussianity must be chosen, in order to discriminate between non-Gaussian and quasi-Gaussian behavior.
 \item[{(iii)}]The maximum discrepancy $\epsilon$ admissible between the truncated series and the LDF must be chosen.
 \item[{(iv)}]A truncated expression is computed with the first n terms.
 \item[{(v)}]At this point, $|t_n|$ provides an estimate of the error. If $|t_n|/|1+\sum_n| \grsim \epsilon$, the error is too large, i.e. the truncated series is not a good approximation to the LDF: a further term must be added and the process resumed at step {(iv)}. This step might require computing further cumulants, as higher-order terms of the series include progressively higher-order cumulants.
 \item[{(vi)}]As the number of terms retained increases, $|t_n|$ becomes smaller than $\epsilon$ and the expression progressively approaches the LDF, until it eventually becomes an acceptable approximation. At this point, if $|\Sigma_n| < \delta$ the pLDF is quasi-Gaussian; otherwise, it is strongly non-Gaussian, a fact that must be taken into account when the distribution is analyzed. In either case, the approximated expression can be used.  

\end{itemize}

\noindent The algorithm is summarized in the flux diagram of Fig.~\ref{fig:edgeworth1}, which permits to find an explicit analytical expression 
for a pLDF of unknown shape but known cumulants.

As an example, we give here the explicit expression of the Edgeworth's series
truncated to include terms up to $n=2$:

\begin{eqnarray}
 \varphi_{\mathrm{L_{tot}}}(x) &= &\frac{1}{\sqrt{2\pi}} e^{-\frac{1}{2} x^2} \times \nonumber \\
  && \biggr( 1 + \frac{1}{6}\, \Gamma_1 \, (x^3 -3 x) + \frac{1}{24}\,\Gamma_2 \,(x^4 - 6x^2 +3)+ \nonumber \\
  && \frac{1}{72}\, \Gamma_1^2 \,(x^6-15 x^4 +45 x^2 -15)\biggl).\label{eq:edge}
\end{eqnarray}

\begin{figure*}
\resizebox{\hsize}{!}{\includegraphics{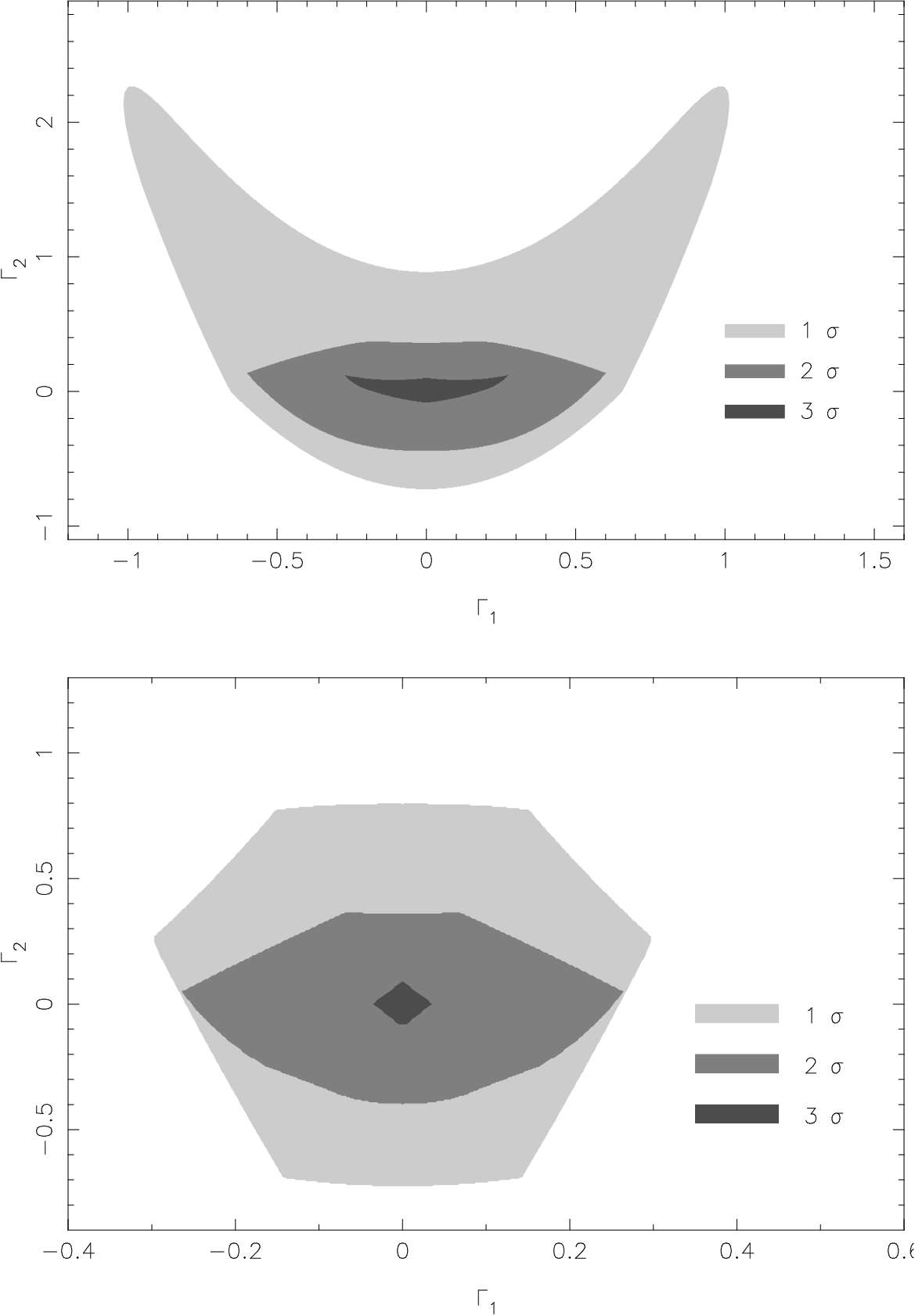} angle=180}
\caption[]{Top: $\Gamma_1$ and $\Gamma_2$ values for which the Edgeworth's expression truncated to $n=2$ approximates the pLDF with a 10\% accuracy or better (Eq.~\ref{eq:edge}). Different shades correspond to different ranges in normalized luminosity.
Bottom: $\Gamma_1$ and $\Gamma_2$ values for which the pLDF can be approximated by a Gaussian to better than 10\%, within different ranges in normalized luminosity.}
\label{fig:smile}
\end{figure*}

\noindent 
In the top panel of Fig.~\ref{fig:smile} we represent the 
region where Eq.~\ref{eq:edge} satisfies the first test of Fig.~\ref{fig:edgeworth1}, that is the range of 
$\Gamma_1$ and $\Gamma_2$ values in which Eq.~\ref{eq:edge}
approximates the pLDF with an accuracy of 10\% or better in a given 
interval of normalized luminosity of $x$. The
dependence on $x$ is represented in Fig.~\ref{fig:smile}
by different shades of gray.

Similarly, the bottom panel of Fig.~\ref{fig:smile} describes the 
region satisfying the second test, i.e. the region 
where the pLDF can be approximated by a Gaussian within a 10\%.
As expected, this region is centered around the point [$\Gamma_1=0, \Gamma_2=0$],
where a true Gaussian would lie. 
Again, this depends on the range of $x$ considered: the wider this range, 
the narrower the range of acceptable $\Gamma_1$ and $\Gamma_2$ values. 

\begin{figure}
\resizebox{8.5cm}{!}{\includegraphics{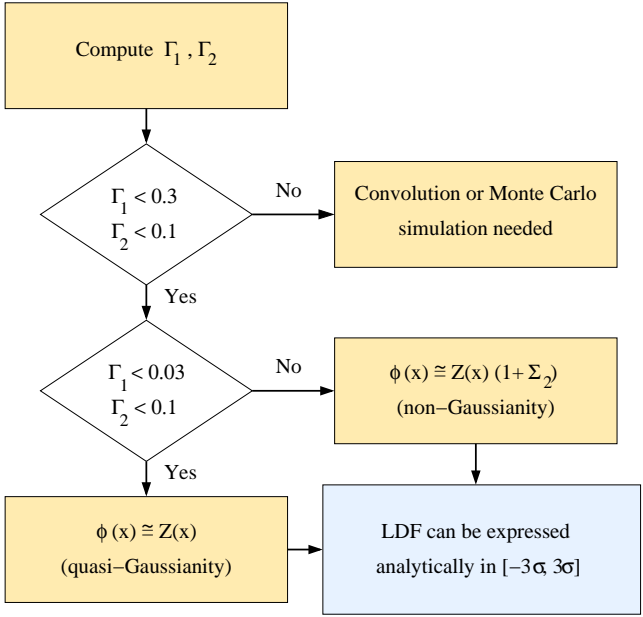}}
\caption[]{Characterization of a pLDF based on Edgeworth's approximation to the second order and a Gaussianity tolerance interval of $\pm$ 10\%.}
\label{fig:edgeworth2}
\end{figure}

The extension of the dark-gray region can be used to define a simplified diagnostic test for luminosity functions (Fig.~\ref{fig:edgeworth2}). 
This test is based on the algorithm of Fig.~\ref{fig:edgeworth1},
but only the first four moments are used, and conventional figures of 10\% are used to define whether Edgeworth's approximation is acceptable and the LDF is quasi-Gaussian. Since $\Gamma_1$ and $\Gamma_2$ are relatable to the skewness and kurtosis of the sLDF via $N_\mathrm{tot}$ (Eqs.~\ref{eq:g1_G1} and \ref{eq:g2_G2}), this test can also be used to determine the minimum number of stars necessary to ensure Gaussianity in a given band. An example of this technique will be given in Sect.~\ref{sub:char}.

\subsubsection{Computation of the variance of the pLDF
}\label{sub:variance}

We have seen in Sect.~\ref{sub:binning} that the Poisson distribution
is only an approximation
to the distribution of the number of stars in a bin, and that it is not a safe one.
The correct alternative is to use the
multinomial distribution, which has the additional advantage of keeping 
track of covariance effects across bins.
However, the multinomial distribution by itself does not tell the whole story.
Additionally, its handling is difficult.
We will show here that the adoption of the 
probabilistic
formulation
permits bypassing this problem, because it provides the tools to compute the relevant quantities without
the need of making assumptions on the distribution of stars in bins,
except its randomness\footnote{This 
is by itself, of course, a strong physical assumption.
If, for example, mass segregation affects star formation, then
this assumption would be false. In that case, the whole problem should
be restated, since the mechanism of star formation should be
incorporated in the Star Formation History.}.

As an example, let's use our 
probabilistic
formalism to compute the variance of 
the distribution. To do so, we need not make any assumption on the distribution
of the number of stars in bins: we just
need to take into account the scale relations
of
luminosity functions (Eq.~\ref{eq:scale})
and the properties of cumulants (Eqs.~\ref{eq:kappa1}$-$\ref{eq:kappa4}):

\begin{equation}
M_2 \, = \, K_2 \, = \, N_{\rm tot}  \, \times \, \kappa_2 = \, N_{\rm tot}  \, \times \, \mu_2. 
\end{equation}

\noindent If we apply the approximation of Eq.~\ref{eq:muwi}, we obtain: 

\begin{eqnarray}
M_2 &=&N_{\rm tot}  \, \times \,  \biggr( \int_0^\infty \ell^2 \, \varphi_{\mathrm{L}}(\ell) \, d\ell \,- \, \biggr( \int_0^\infty \ell \, \varphi_{\mathrm{L}}(\ell) \, d\ell \biggl)^2\biggl) = \nonumber \\
&=& N_{\rm tot}  \, \times \,  \biggr(\sum_i  w_i \,\ell_i^2 \, -\,  \biggr(\sum_i  w_i \,\ell_i \biggl)^2 \biggl)= \nonumber \\
&=& N_{\rm tot}  \, \times  \,\biggr(\sum_i w_i \ell_i^2 - \sum_i (w_i \ell_i)^2 -  \sum_i \sum_{j \neq i} \ell_i \ell_j \, w_i \, w_j\biggl),
\label{eq:varwiphi}
\end{eqnarray}

\noindent which is the same as the expression of Eq.~\ref{eq:varwiM}.
This simple example shows the power and, at the same time, the simplicity
of our 
probabilistic
formalism. 
Note, however, that this result does not imply that the multinomial description is wrong: on the contrary, it is implicit in the 
probabilistic
treatment. Our point here is that the 
probabilistic
treatment is simpler and more powerful than an analysis explicitly based on the multinomial distribution.

Finally, note that the value of the 
variance of the pLDF
customarily assumed in the literature (Eq.~\ref{eq:varwiP}) is biased, 
with the work by \cite{CRBC03} as the only exception we are aware of. 
Note that in fact, the variance obtained from Eq. \ref{eq:varwiP} is 
the second  raw moment of the 
pLDF,
and not its second central moment. 

\subsection{Technical problems in the computation of the 
pLDF}\label{sub:techproblems}

Unfortunately, the determination of the LDF of an ensemble is hindered by a few technical problems. 
The first one concerns the stellar LDF: although some aspects of stellar populations (like variability or transient events) would be more easily treated via the luminosity function as compared to the standard method, the 
modeling of the luminosity evolution during
fast evolutionary phases poses the same problems as in standard codes. 

A second issue is the contribution of dead stars to the luminosity function. Those stars have no influence on the computation of the moments of the distribution, except for a ``normalization factor'' that depends on the age of the population.
In terms of the shape of the luminosity distribution, dead stars show up as a  pulse function located at zero luminosity.
This
can be easily understood in the following terms: let us assume an ensemble with
$N_\mathrm{tot}$ initial stars. At a given age 
there is a non-zero probability that all stars are dead; 
the 
pLDF,
marginalized to such a condition, is a delta centered on zero. There is also a non-zero probability that all but 1 stars are dead: the 
pLDF
corresponding to this case is the sum of a pulse function and the stellar luminosity function, the pulse function having in this case a strength smaller than the delta. Similarly, all but 2 stars can be dead, and the corresponding 
pLDF
is a pulse function plus the convolution of two stellar luminosity functions. In total, there are $N_\mathrm{tot}$ + 1 marginalized cases, corresponding to $N_\mathrm{tot}$, $N_\mathrm{tot} -1$, $N_\mathrm{tot} -2$, ..., 2 and 1 dead stars.
The LDF of the ensemble should include all these cases, each with 
a weight given by its relative probability, so that it will have a pulse contribution centered on zero. Although this result seems to contradict the central limit theorem, which states that the asymptotic shape of the LDF of the ensemble is a Gaussian, this is not the case, since the relative pulse contribution  becomes smaller and smaller as $N_\mathrm{tot}$ increases.

\section{Applications of the probabilistic treatment}\label{sect:application}

\subsection{Characterization of a sLDF
by means of its cumulants}\label{sub:char}

\begin{figure*}
\resizebox{\hsize}{!}{\includegraphics{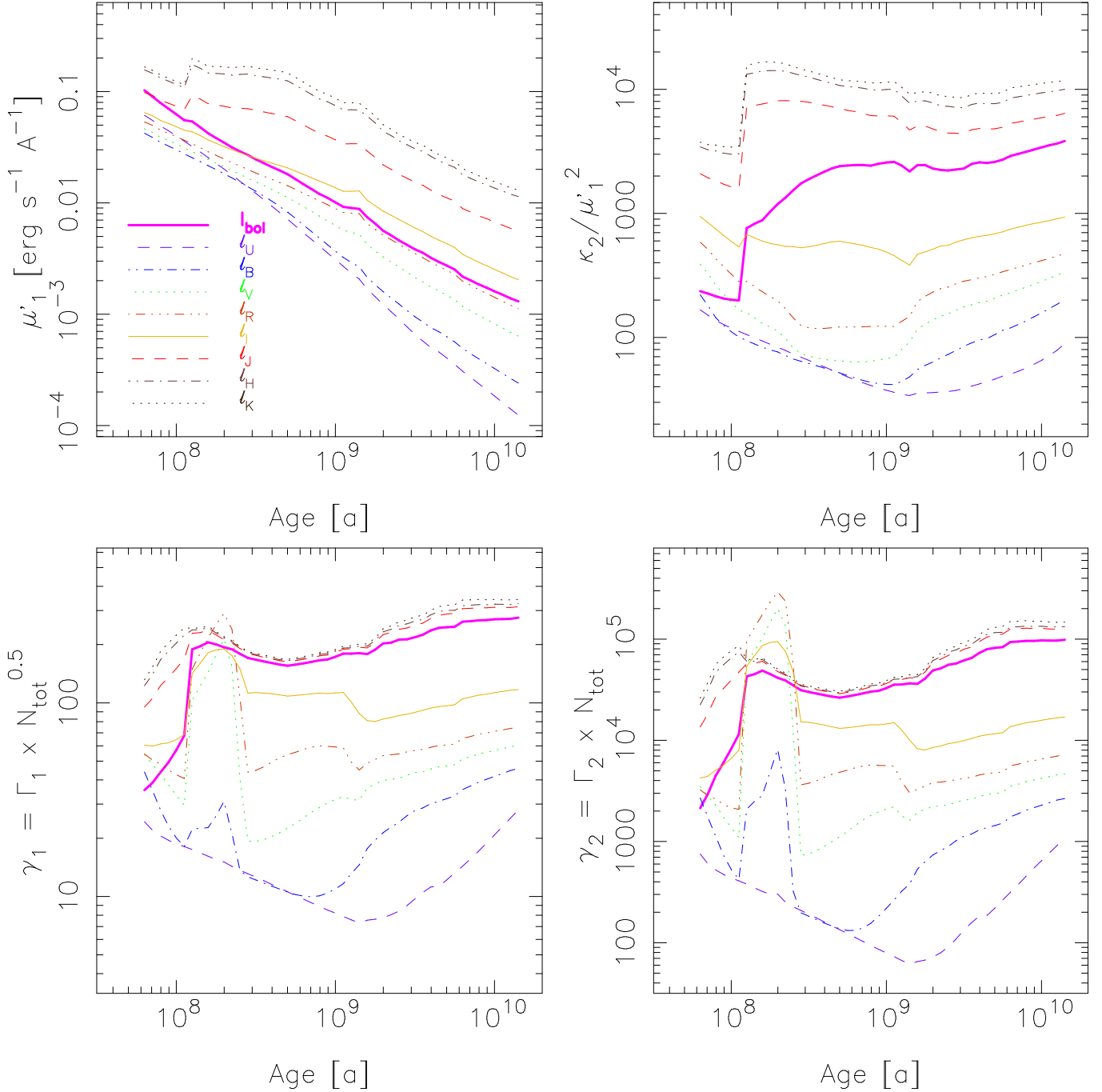}}
\caption[]{Main parameters
of the luminosity function in several photometric bands, obtained from the isochrones by \cite{MG01} and adopting the IMF by \cite{Sal55} in the mass range 0.15 - 120 M$_\odot$.}
\label{fig:cum}
\end{figure*}

In Sect.~\ref{sub:lmax} we showed that knowledge of the mean and the variance is not enough to characterize a LDF. As a rule of thumb, at least 
the first four cumulants should be taken into consideration, which describe the mean, the dispersion, the asymmetry and the peakedness of the distribution. As a first example of application of the 
probabilistic
treatment, we 
will derive a few properties of stellar population from the analysis of its cumulants.
Fig. \ref{fig:cum} shows the first four cumulants of the 
sLDF
for different bands and ages, 
obtained from
the isochrones by \cite{MG01}. We have assumed a \cite{Sal55} IMF in the mass range 0.15 - 120 M$_\odot$ normalized by the total number of stars. For a normalization in mass rather than in number of stars,
these results must be multiplied by the mean mass, $\langle m \rangle = 0.52$ M$_\odot$.
The figure illustrates
several interesting facts: 

\begin{itemize}

\item The value of $\kappa_2/\mu_1'^2=\mu_2/\mu_1'^2$ is in general larger than 10, with the exception of blue bands. As has been pointed out in Sect.~\ref{sub:variance},
most synthesis models compute 
the second raw moment $\mu'_2$ instead of the variance $\mu_2$. The difference between the two moments is a term in $\mu_1'^2$, but since $\mu_2$ is in general much larger than $\mu_1'^2$, 
$\mu_2'$ is numerically close to $\mu_2$. 
Although the variance computed by most synthesis codes is systematically biased, in numerical terms the value is nearly correct.
However, it is safer to compute the variance properly.

\item As pointed out in previous works, the Gaussian regime is not as common in stellar populations as often assumed. This is particularly the case of infrared wavelengths, whose values of $\gamma_1$ and $\gamma_2$ imply, given the criterion of Fig.~\ref{fig:edgeworth2}, that
at least $10^7$ stars are needed
(roughly $\approx 5 \times 10^6$ M$_\odot$ in initial mass) in the analysis building block (i.e. individual pixels in SBF studies, or the whole slit in integrated data) to ensure quasi-gaussianity. Non-gaussianity 
is a problem when one wants to obtain confidence intervals in terms of $\sigma$:  when the distribution is non-Gaussian, it cannot be assumed that $\mu_1' \pm 1 \sigma$ contains $\approx 68$\% of the distribution.
This is not necessarily a problem for goodness-of-fit tests like $\chi^2$,  
since the test also works for non-Gaussian distributions provided the deviation from 
gaussianity is not severe.
For example, $\Gamma_1$ should be used to have an approximate idea of which test would be the best to compare the observations with the models.

\end{itemize}

\subsection{Inclusion of uncertainties in the input parameters}\label{sub:uncertainties}

The input ingredients used in population synthesis are generally assumed to be fully known but are in fact affected by uncertainties \citep{CL05}. 
These uncertainties may either reflect incomplete knowledge or an intrinsic spread in the features of the quantities considered.
In either case, they can be included in the sLDF provided they can be modeled quantitatively.

As an example, let's suppose that the mass distribution function depends on a parameter $\theta$ that is itself distributed, that is let's replace the univocally determined function $\varphi_m(m)$ with a parametric function $\psi(m;\theta)$. The parameter $\theta$ can be characterized by a probability distribution function such that:

\begin{equation}
\int p(\theta) \, d\theta = 1,
\end{equation}

\noindent where the integration interval is the range where $\theta$ is defined.
To apply Eq.~\ref{eq:ldf-imf}, it is necessary to eliminate the parametric dependence. This is done by integrating $\psi(m;\theta)$ over all possible $\theta$ values, weighting it by $p(\theta)$:

\begin{equation}
\varphi_m(m) = \int \psi(m;\theta) p(\theta) \, d\theta,
\end{equation}

\noindent The resulting distribution, which is technically called a {\it mixture} of $p(\theta)$ and $\psi(m;\theta)$ \citep{KS77}, can be used in Eq.~\ref{eq:ldf-imf}.

Let's consider two hypothetical examples. In the first, assume that $\psi(\theta;m) \propto m^{-\theta}$ and that the power-law index $\theta$ has a
rectangular distribution centered on Salpeter's index $\alpha=2.35$:

\begin{equation}
p(\theta) = \Biggl\{
 \begin{array}{lr}
         \frac{1}{2\delta} & \hfill \mathrm{if}\; - \delta \le \theta -2.35 \le + \delta,\\
          0 & \mathrm{otherwise},\\
 \end{array}
\end{equation}

\noindent with $\delta>0$.
If we mix $\psi(\theta;m)$ with $p(\theta)$ and integrate, we find:

\begin{eqnarray}
\varphi_m(m)&=&\int \psi(\theta;m) \, p(\theta)\, d\theta \propto \nonumber\\
            &\propto& \frac{(m^{\delta}-m^{-\delta})}{2\delta\, \mathrm{ln}\, m} \, m^{-2.35}.
\end{eqnarray}

\begin{figure}
\resizebox{\hsize}{!}{\includegraphics{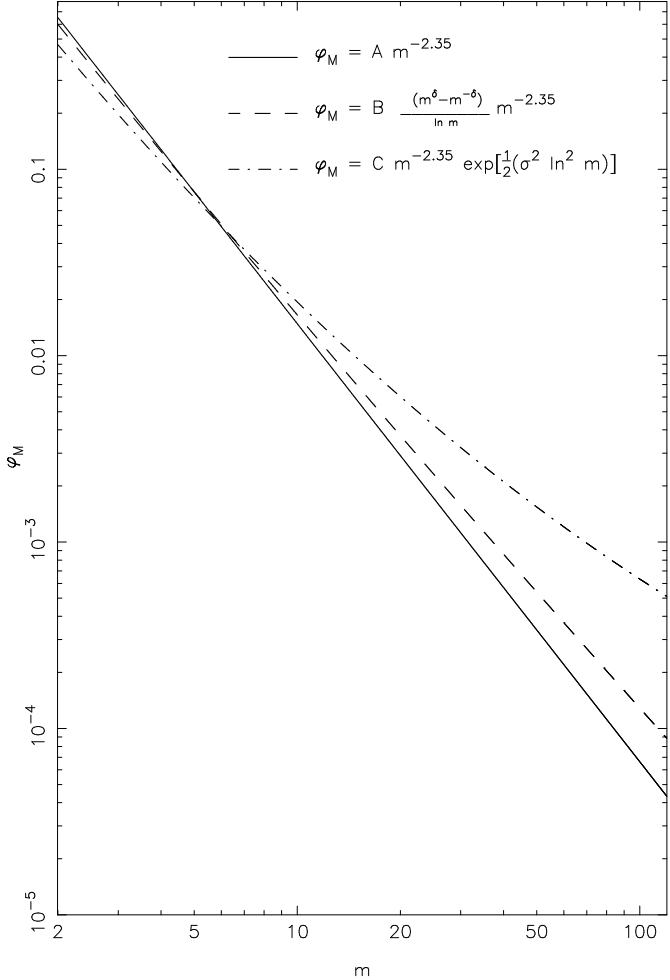}}
\caption[]{Salpeter's IMF (solid line) compared to two examples of distributed IMFs: rectangular (dashed, $\delta=0.5$) and Gaussian (dot-dashed, $\sigma=0.5$).}
\label{fig:IMFdistr}
\end{figure}

\noindent When $\delta \rightarrow 0$, the expression ${(m^{\delta}-m^{-\delta})}/{(2 \delta\, \mathrm{ln}\, m)} \rightarrow 1$ and Salpeter's law is recovered. When $\delta$ is non-negligible, the sLDF is distorted with respect to the simple Salpeter's case
(Fig.~\ref{fig:IMFdistr}).

As a further example, let's assume that $\psi(\theta;m) \propto m^{-\theta}$ as above, 
but that now $p(\theta)$ is gaussian:

\begin{equation}
p(\theta) =  \frac{1}{\sigma_\theta \sqrt{2\pi}} e^{-\frac{(\theta - 2.35)^2}{2\sigma_\theta^2}}.
\end{equation}

\noindent In this case, the $\theta$-weighted IMF is given by $\varphi_m(m)\propto m^{-2.35} exp(\sigma_\theta^2 ln^2 m)/2$ and, again, the IMF is distorted with respect to the limiting case given by Salpeter's law (Fig.~\ref{fig:IMFdistr}).

Following this example, any spread in the input ingredients can be included in the modeling and contribute to the shape of the pLDF. 
In particular, the method outlined above can be used to include
transient phases and fuzzy stellar behavior in the modeling (Sect.~\ref{sub:trans}).
For this to be done, it is just necessary that the phenomenon considered can be 
described in terms of a parameter of known distribution function. 
Other uncertainties that can be incorporated in the modeling are those 
that reflect our imperfect knowledge of the problem: for example, the sLDF could be mixed with a Gaussian distribution to mimic observational errors.

It is important to note that the effect on the pLDF of a spread in the input ingredients cannot be determined {\it a priori}, as it depends on how the distribution of the input ingredient affects the sLDF. In particular, it cannot be established whether the inclusion of distributed ingredients will render convergence to gaussianity more or less rapid. It is to be expected that, the further is the sLDF from gaussianity, the slower will the convergence of the pLDF to gaussianity be. For example, 
if the distribution of the input ingredient is bimodal, bimodality will persist in the pLDF of small $N_\mathrm{tot}$, and a larger number of stars will be required for the pLDF to converge to a Gaussian.

\subsection{Comparison between Monte Carlo simulations, numerical convolutions and the Edgeworth's approximation}

\begin{figure}
\begin{center}
\resizebox{8cm}{!}{\includegraphics{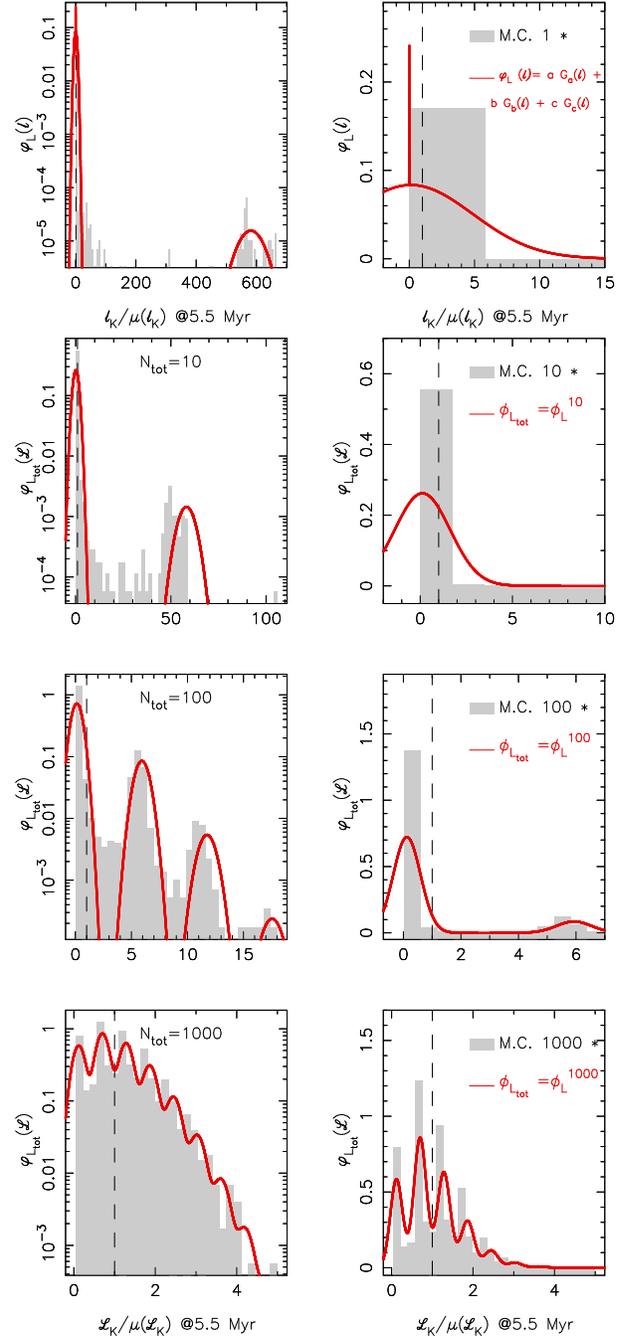}}
\end{center}
\caption[]{
Monte Carlo simulations of clusters at solar metallicity and 5.5 Ma in the K band  for different values of $N_\mathrm{tot}$ 
 \citep[][shaded histograms]{CVG03},
compared to analytical pLDFs obtained 
convolving $N_\mathrm{tot}$ times a sLDF made up of three Gaussians (solid line).
The vertical scale is logarithmic in the left panels and linear in the right panels. 
The analytical sLDF has the same mean and variance of the Monte Carlo simulations;
the position of the mean is shown by a vertical dashed vertical line.}
\label{fig:MC}
\end{figure}

We have shown in Sect.~\ref{sub:exact_pLDF} that 
obtaining the pLDF through a convolution of the
sLDF is not simple; however, in simple scenarios
it is possible to solve it. In the following, we will consider a simple case
to illustrate this point. Our aim in this experiment is twofold:
first, we will derive an explicit expression for the pLDF at different $N_\mathrm{tot}$'s
to show how its shape is related to the generating sLDF
and how it depends on $N_\mathrm{tot}$.
Second, we want to show that, through our method, we are able to
reproduce the main features of Monte Carlo simulations for any value of $N_\mathrm{tot}$.

Let us assume a stellar luminosity function made up of three Gaussians, representing the dead stars, the MS, and the PMS respectively. The parameters of the three Gaussians are chosen so that the broad features of a set of Monte Carlo simulations for one star are reproduced (upper panels of Fig.~\ref{fig:MC}; the vertical scale is logarithmic in the left panel and linear in the right panel). In particular, the mean and the variance of the triple-Gaussian LDF are constrained to be the same as those of the Monte Carlo simulations. We also confirmed {\it a posteriori} that $\gamma_1$ and $\gamma_2$ are very close, which is expected for similar distributions. 

With a numerical routine, we convolved this sLDF with itself $N_\mathrm{tot}$ times. The resulting pLDFs for selected values of $N_\mathrm{tot}$ are shown in Fig.~\ref{fig:MC} (solid line). 
The features of these pLDFs can be understood qualitatively as follows: 
the characteristic function of this 
sLDF
is:

\begin{eqnarray}
\phi_\mathrm{L}(p) =  &A_\mathrm{ds} e^{-\frac{1}{2} \sigma_\mathrm{ds}^2 p^2 - i \ell_\mathrm{ds} p}  
+ A_\mathrm{MS} e^{-\frac{1}{2} \sigma_\mathrm{MS}^2 p^2 - i \ell_\mathrm{MS} p}  + & \nonumber \\
& + A_\mathrm{PMS} e^{-\frac{1}{2} \sigma_\mathrm{PMS}^2 p^2 - i \ell_\mathrm{PMS} p},&
\end{eqnarray}

\noindent where $A_\mathrm{ds}$, $A_\mathrm{MS}$, and $A_\mathrm{PMS}$ are the weights of the Gaussians corresponding to dead stars, the MS, and the PMS respectively; $\ell_\mathrm{ds}$, $\ell_\mathrm{MS}$, and $\ell_\mathrm{PMS}$ their locations on the luminosity axis; and $\sigma_\mathrm{ds}$, $\sigma_\mathrm{MS}$, and $\sigma_\mathrm{PMS}$ the respective dispersions.
The exponent of the $N_\mathrm{tot}$-th power of this characteristic function is a sum of real exponents in $p^2$ (i.e. Gaussian distributions) and imaginary exponents in $p$
(i.e. translations of the corresponding distributions). Hence the final function will be a sum of Gaussians located at different positions.

The vertical sequence of panels shows how, increasing $N_\mathrm{tot}$, the pLDF progressively becomes smoother and more symmetric, approaching a Gaussian shape. 
In the same figure, Monte Carlo simulations with corresponding $N_\mathrm{tot}$ values are also shown \citep{CVG03}, and these coincide remarkably with the analytical pLDF. This is a consequence of having chosen a sLDF similar to the Monte Carlo distribution function for $N_\mathrm{tot}=1$ (which, in turn, maps the underlying sLDF), and shows the power of the method: large Monte Carlo simulations become redundant, in the sense of being predictable, if one can characterize the sLDF and succeeds in convolving it.

\begin{figure}
\begin{center}
\resizebox{8cm}{!}{\includegraphics{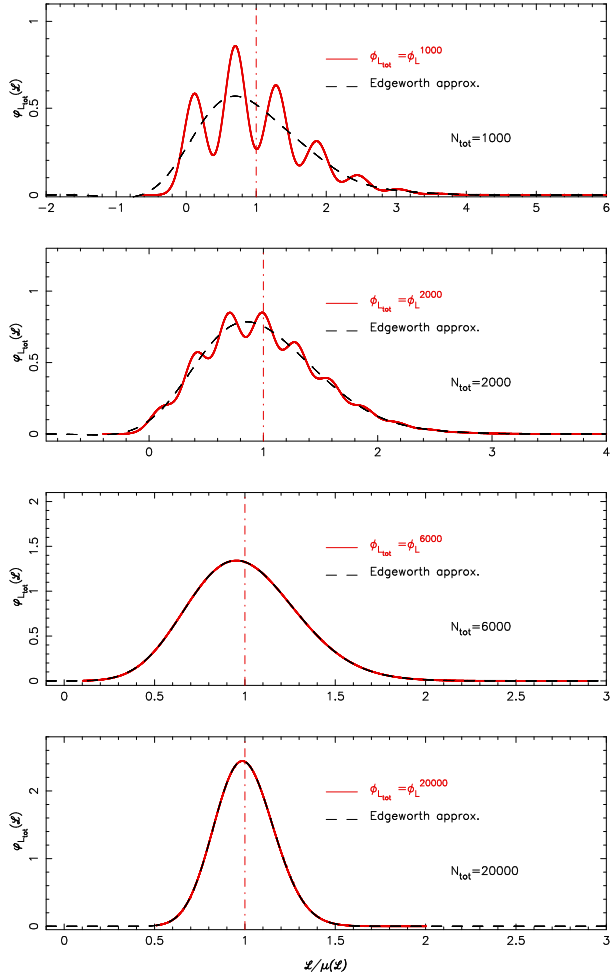}}
\end{center}
\caption[]{Comparison between the pLDFs obtained through sLDF convolutions (dashed line) and the Edgeworth's approximation (solid line) for clusters with different $N_\mathrm{tot}$ values.}
\label{fig:edgeconv}
\end{figure}

Fig.~\ref{fig:edgeconv} compares the pLDFs obtained through convolution of the sLDF to their Edgeworth's approximation to the second order (Eq.~\ref{eq:edge}). Each panel corresponds to a different number of stars; the cluster in the upper panel, with $N_\mathrm{tot}=1000$, is the same one of the lower panel of Fig.~\ref{fig:MC}. The Edgeworth's approximation improves visibly across the range of $N_\mathrm{tot}$ considered: this is expected, because the closer is the pLDF to a Gaussian, the better is it approximated by Edgeworth's expression.

\begin{figure}
\begin{center}
\resizebox{8cm}{!}{\includegraphics{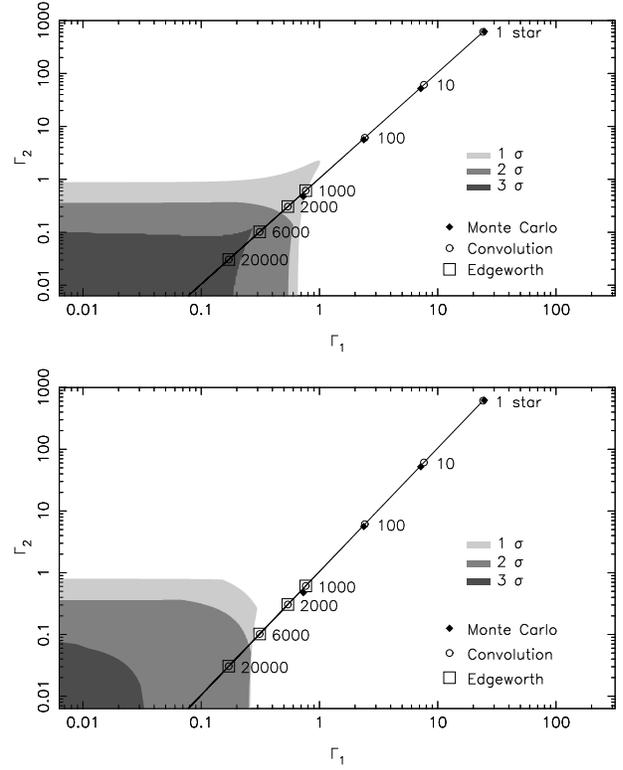}}
\end{center}
\caption[]{$\Gamma_1$ and $\Gamma_2$ values of the pLDFs obtained through Monte Carlos simulations (black diamonds), sLDF convolutions (empty circles), and the Edgeworth's approximation (empty squares). Larger $\Gamma_1$ and $\Gamma_2$ correspond to lower $N_\mathrm{tot}$ values.}
\label{fig:smileMC}
\end{figure}

Finally, in Fig.~\ref{fig:smileMC} we show a blow-up of Fig.~\ref{fig:smile} that compares the $\Gamma$s of the pLDF obtained through the three different methods (Monte Carlo simulations, numerical convolution, and the Edgeworth's approximation). Several points deserve to be emphasized here:

\begin{itemize}

\item As noted above, the $\Gamma_1$ and $\Gamma_2$ values of the Monte Carlo simulations
nearly coincide with those of the convoluted pLDF. This is a remarkable proof of the power
of describing stellar populations in terms of luminosity distributions.

\item The dots corresponding to clusters of 10$^3$ stars fall within the shaded region in the top plot (approximation test) but not in the bottom one (Gaussianity test), implying that the pLDF can be approximated by the Edgeworth's function, but is still far from Gaussianity. This is also apparent from the bottom panel of Fig.~\ref{fig:MC}, which shows that the shape of the pLDF (described both as a convolution and by Monte Carlo simulations) is markedly multimodal and asymmetric, hence far from Gaussian.

\item Finally, in this example Gaussianity is marginally reached only above $N_\mathrm{tot}=6\cdot10^3$ stars, and even a cluster as large as  $N_\mathrm{tot}=2\cdot10^4$ is Gaussian only within the $[-2\sigma, 2\sigma]$ interval.
\end{itemize}

\subsection{Criteria for assessing the 
significance
of fits}

As has been pointed out several times, the main result of synthesis models is the mean value of the stellar luminosity function, which can be scaled to clusters of any size. We have also shown that the relative dispersion of the model results (the ratio $\sigma({\cal L})/M_1' = \sqrt{M_2}/M_1'$) decreases when $N_\mathrm{tot}$ increases. However, $\sigma({\cal L})$ increases in absolute terms, since it is proportional to $\sqrt{N_\mathrm{tot}}$, 
a fact that
should be taken into account in the comparison of theoretical data to observed clusters. Furthermore, since each monochromatic luminosity has its own $\sigma({\cal L})$, 
they should be weighted differently in fits.
Finally, although some regions of the spectrum may have a quasi-Gaussian distribution, this will not happen in general with all the regions. Hence, not all the 
frequencies (either in synthetic or in observed spectra)
are equivalent, or even suitable, to obtain the properties of the observed cluster.

\begin{figure}
\resizebox{\hsize}{!}{\includegraphics{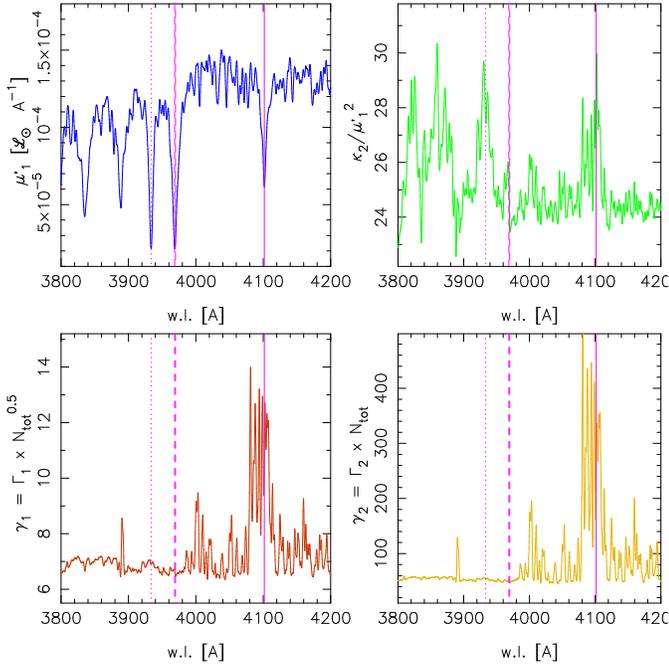}}
\caption[]{Cumulants of the monochromatic spectral energy distribution for a 1 Ga luminosity function assuming a \cite{Sal55} IMF in the mass range 0.15 - 120 M$_\odot$. The figure shows the region around H$\delta$, which is marked with a solid line. The lines Ca {\sc ii} H + H$\epsilon$  are marked with dashed lines, and the line Ca {\sc ii} K with dotted lines.}
\label{fig:SED}
\end{figure} 

As an example, Fig. \ref{fig:SED} shows the first four cumulants of a region of the visual electromagnetic spectrum for the 
pLDF 
of a 1 Ga cluster with solar metallicity, obtained from the synthesis code sed@\footnote{sed@ is a
synthesis code included in the {\it Spanish Virtual Observatory} and the {\it Violent Star Formation Legacy Tool project} by means of the {\it PGos3} tool. 
 The code is written in ANSI C under GNU Public License and, currently, is managed by M. Cervi\~no; use of the code and its results must be referred to solely by its documentation ({\it Sed@ Reference Manual},
in preparation), its WWW address and the citations in the headers of the output VO Tables. The code is currently accessible on-line at 
{\tt http://ov.inaoep.mx/}. Its inclusion in the VO service grid is under way.} using a  
\cite{Sal55} IMF in the mass range 0.1-120 $M_\odot$. The results are normalized to mass.
The isochrones used are from \cite{Gi02} covering a mass range from 0.15 to 100 M$_\odot$ and based on the solar models (Z=0.019) by \cite{Gi00} and \cite{Be94} that include overshooting and a simple synthetic treatment of the thermal pulses AGB phase \citep{GB98}. The atmosphere models are taken from the high resolution library by \cite{Metal04,GDetal04} based on PHOENIX \citep{1999JCAP..102...41H,2001ApJ...556..357A}, the ATLAS 9 odfnew library \citep{2003IAUS..210..A20C} computed with SPECTRUM \citep{1994AJ....107..742G}, the
ATLAS 9 library \citep{1991sabc.conf..441K} computed with SYNTSPEC \citep{HLJ95}, and TLUSTY \citep{2003ApJS..146..417L} at [Fe/H] = 0.0 dex.

The figure shows the region around H$\delta$. As expected from previous plots at these ages (Fig. \ref{fig:cum}), the values of $\Gamma_1$ and $\Gamma_2$ are quite low. However, not all the wavelengths have the same statistical significance. In particular, the H$\delta$ line shows the largest values of $\Gamma_1$ and $\Gamma_2$, so in undersampled clusters its profile will be difficult to fit. On the other hand, the profile of Ca {\sc ii} H + H$\epsilon$ is a quite robust result, with a low relative dispersion and $\Gamma_1$ and $\Gamma_2$ values close to the continuum level. Finally, the Ca {\sc ii} K line, with $\Gamma_1$ and $\Gamma_2$ values similar to those of the continuum, has a high relative dispersion. Summarizing, fitting the theoretical models of Ca {\sc ii} lines, including their profiles, to observed data would yield more realistic results than fitting either the intensity, the equivalent width or the profile of H$\delta$. Of course, these conclusions 
depend on the age and metallicity.

\section{Future applications of the probabilistic treatment}\label{sect:future}

In this section we will briefly discuss several potential extensions of the formalism that could have a strong impact in the analysis of stellar populations. Details on a 
a few examples will also be given.

\subsection{Forthcoming extensions of the formalism}\label{sub:forthcoming}

Since in this paper we have only considered the case of SSPs, maybe the most important 
pending issue is the extension to other scenarios of star formation. This topic will be
covered in a forthcoming paper.

A current limitation of the formalism is that it only deals with integrated properties that scale linearly with the number of stars in the cluster. In the future, the formalism will be extended to include the case of luminosity ratios.
To solve this problem,
in addition to  the cumulants of the distribution function of the ensemble, it is also necessary to obtain the correlation function of the corresponding quantities.

A further assumption of the formalism in its present stage is that the stellar population 
has a fixed
number of stars $N_\mathrm{tot}$. 
It is our intention to extend the formalism by including
the case of a collection of populations with varying number of stars.
This 
extended formalism could be applied to several problems, 
such as 
the analysis of stellar populations in pixels,
which requires computing
the global distribution resulting from the distribution of stars in each pixel and the distribution of numbers of stars across pixels; 
the distribution of luminosities in globular clusters; the estimation
of the difference between 
luminosity profiles in galaxies inferred by means of a comparison with the mode and the mean respectively; 
the comparison of theoretical SBF with observed ones;
and the comparison among different Monte Carlo simulations performed with a total fixed mass. 
A few of these prospective applications will be discussed further in the remainder of this section.

Finally, the formalism developed here for the treatment of luminosity functions can also be applied to stellar yields. All the statistical considerations that we have made about luminosities can be easily translated into equivalent issues in the field of chemical evolution, 
although in that case
the star formation history would have a fundamental role, and the differential equations that describe (deterministically) the chemical evolution would become stochastic differential equations, whose mean would coincide with the results obtained deterministically. As in the present case, comparing the mean against observations can produce a bias that depends strongly on the shape of the distribution.

The following sections will give more details on a few examples among these.

\subsection{Surface Brightness Fluctuations}

SBF observations from galaxies and globular clusters have been proposed as a test of evolutionary tracks and isochrones \citep[e.g.][ among others]{CRBC03}. 
This test is based on the comparison between the observed variance across pixels in the image of a galaxy and the variance expected on statistical grounds.
However, 
there
are several inconsistencies in this method as is applied at present:

\begin{itemize}

\item Observational SBF are the result of an average over an additional distribution, the one that defines the number of stars falling in a given pixel. The theoretical SBF formalism doesn't take this second distribution in consideration. 

\item Each of the building-blocks of the distribution (the pixels) is representative of the integrated luminosity of an ensemble. Although the formalism, in terms of moments, can be scaled to pixels containing any number of stars, the number of stars in a given pixel and the number of pixels with the same number of stars should be known, in order to evaluate the error in the estimation of the SBF due to the finite sampling of the underlying distribution.

\item The theoretical SBF used to date are computed under the implicit assumption of a SSP, whereas the mode of star formation of a real galaxy can be much more complex.

\end{itemize}

This inconsistencies could be overcome by extending the formalism 
so as to include the LDF
of populations with varying number of sources.
This subject will be discussed at length in a forthcoming paper.

\subsection{Putting constraints on the globular clusters' distribution}\label{sub:gclf}

In a similar way, the luminosity distributions of globular clusters 
in galactic halos
could be compared to the corresponding distributions obtained theoretically, 
either in terms of moments or in terms of the explicit shape of the distribution. 
The possibility of obtaining higher-order moments, such as the skewness, helps constraining the distribution, which is necessary to test evolutionary tracks and isochrones. However, there are drawbacks similar to those of the previous case, with the only exception of the assumption on the validity of a SSP, which 
in a galactic halo is probably verified.

To pursue the goals sketched above, it is imperative to know the initial distribution of cluster masses. In the following, we will show qualitatively that our method can also contribute toward this goal, by disclosing a potential source of bias in the use of synthesis models for the determination of the mass distribution of globular clusters. Note, however, that a quantitative conclusion cannot yet be reached.

\begin{figure}
\resizebox{\hsize}{!}{\includegraphics{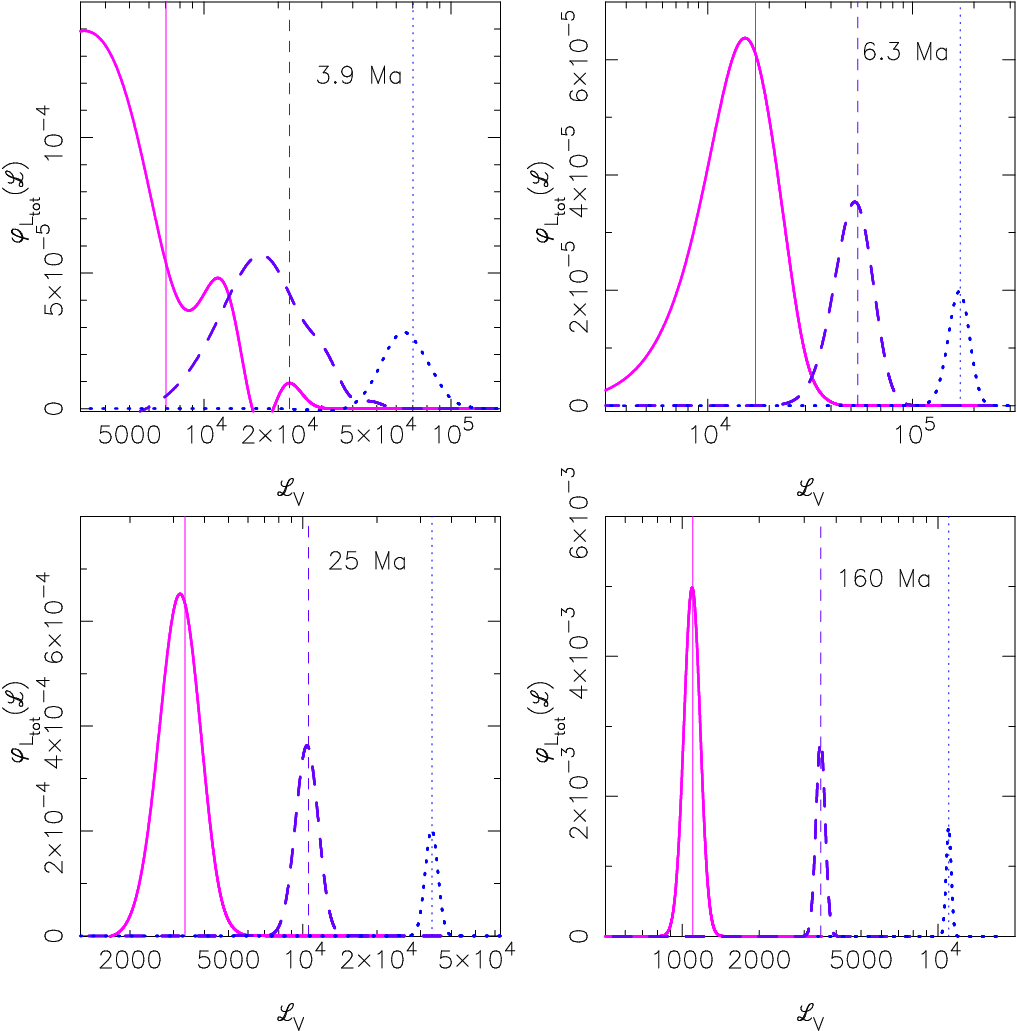}}
\caption[]{Luminosity distribution functions in V for clusters with the LMC metallicity, approximated by the Edgeworth's series, for several ages and cluster masses: 10$^4$ M$_\odot$ (solid line), 10$^{4.5}$ M$_\odot$ (dashed line) and 10$^5$ M$_\odot$ (dotted line). The mean values of the 
pLDFs
are marked as vertical lines.}
\label{fig:ZF}
\end{figure}

In Fig. \ref{fig:ZF} we show the pLDFs
in $\mathcal {L}_\mathrm{V}$ for clusters with the LMC
metallicity obtained from the isochrones by \cite{Gi02} using the Edgeworth's approximation. We have plotted the distributions that correspond to the extremes of the age ranges used by  \cite{ZF99}. The 
pLDFs
correspond to cluster masses of  10$^4$ M$_\odot$ (solid line), 10$^{4.5}$ M$_\odot$ (dashed line), and 10$^5$ M$_\odot$ (dotted line). The mean values of the corresponding pLDFs 
 are marked as vertical lines. 

The figure shows that, given the asymmetry of the stellar luminosity function, most observed clusters will have smaller luminosities than the mean, i.e. smaller luminosities than those predicted by a standard code. The straightforward implication is that the mass of observed clusters inferred by means of a comparison with standard models is, in most cases, underestimated, with a bias larger for smaller clusters 
\citep[see also][]{GGS04}
and younger ages.

 This effect is highly relevant for young and undersampled clusters. In particular, assuming that the age estimation obtained by \cite{ZF99} is not biased, the figure clearly shows that there can be
a systematic  underestimation of the mass of younger clusters, an hence, an overestimation of the number of low-mass clusters. In the cluster mass range considered, this effect is more relevant in the 10$^{4}$ to 10$^{4.5}$ M$_\odot$ interval. 
This fact could
change the shape of the distribution of the initial cluster masses obtained by \cite{ZF99}, particularly in the low mass range, making it shallower or even inverting the slope. However, this result is only a 
qualitative
application of the new formalism, and the example above should not be taken literally. To establish a firm conclusion it is necessary to apply the 
probabilistic
treatment also to the determination of ages.

\subsection{Tracing the sLDF with resolved stellar populations}\label{sub:tracing}

The individual stars in a resolved population can be used to trace the sLDF. This can be done by comparing the first four cumulants computed by means of current synthesis models to the corresponding observed quantities.
Resolved populations have several advantages:

\begin{itemize}

\item  $N_\mathrm{tot}$ is a known quantity (i.e. $N_\mathrm{tot}$=1, except in the case of crowding problems, which would be treated separately).

\item The number of stars used 
to map
the luminosity function and the possible luminosity bias are known,
and can be incorporated theoretically (e.g. by changing the value of $\ell_\mathrm{min}$ in the computation of the moments). 

\item The assumption of a SSP can be directly tested. 
Moments, and their associated sampling errors, can be  estimated from simple star counts using the standard formulations for unbiased estimations of the mean, variance, skewness and kurtosis.

\item It is possible to perform a tailored analysis consistent with the used isochrones without outsiders; e.g. Blue Stragglers can be excluded from the estimation of the moments if they are not included in the theoretical luminosity function. 

\item Finally, the covariance effects between different evolutionary phases are included in a natural way in the 
probabilistic
treatment, both theoretically and observationally. 

\end{itemize}

\begin{figure}
\resizebox{\hsize}{!}{\includegraphics{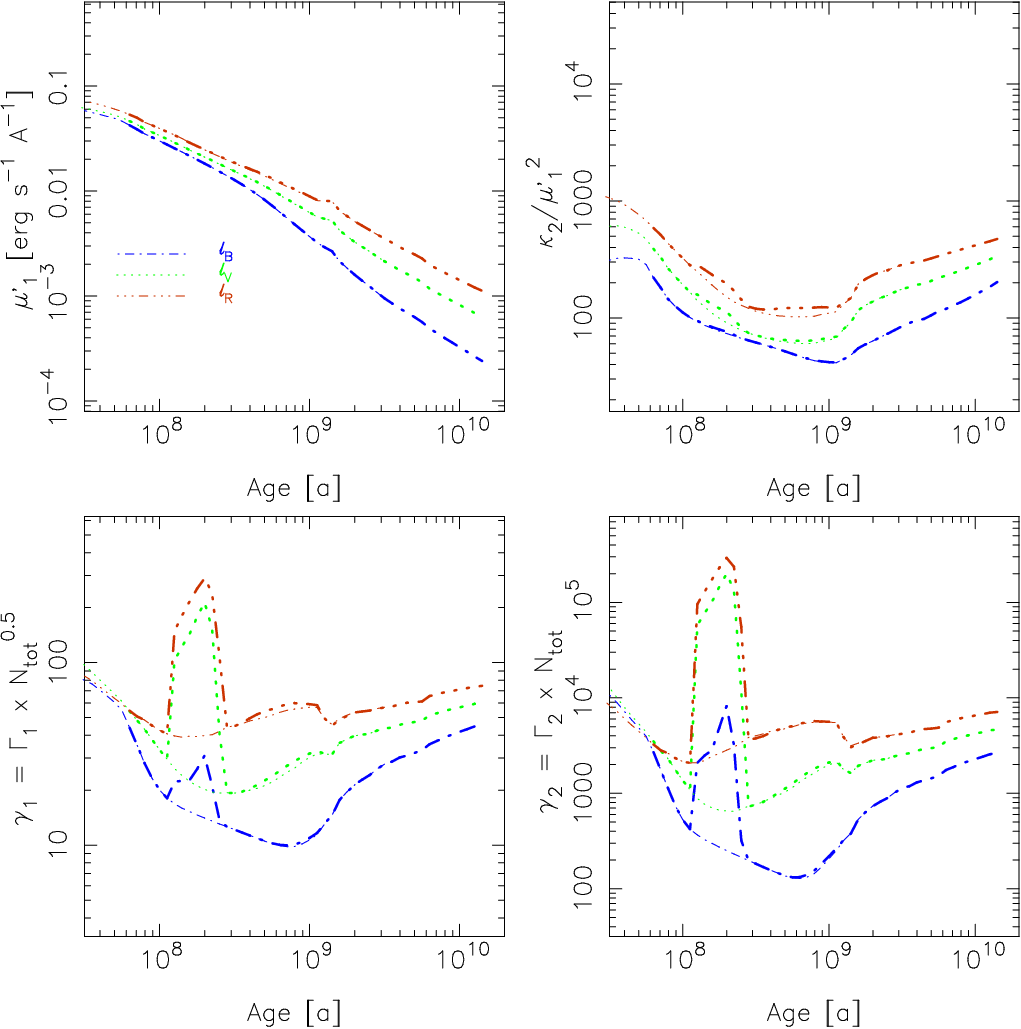}}
\caption[]{Main parameters
of the luminosity function for several photometric bands obtained from the isochrones by \cite{MG01} (bold lines) and \cite{Gi00,Be94} (light lines), assuming a \cite{Sal55} IMF in the mass range 0.15 - 120 M$_\odot$.}
\label{fig:cumiso}
\end{figure} 

\subsection{Testing the isochrones}

The method proposed in Sect.~\ref{sub:tracing} can also be used to compare the predictions of different sets of isochrones, since it is also sensitive to the number of stars expected in each region of the isochrone (which is almost IMF-independent for PMS stars).
A similar method has been proposed by \cite{WH03}.

To illustrate the point,
we show in Fig. \ref{fig:cumiso} the first four cumulants of the 
sLDF
for different bands and ages using the isochrones by \cite{MG01}, computed following FCT requirements explicitly,  and by \cite{Gi00} and \cite{Be94}. We have assumed a \cite{Sal55} IMF in the mass range 0.15 - 120 M$_\odot$ normalized by the total number of stars, and the results have been obtained by a direct integration of the isochrones. 
Note that, although the mean values and the $\mu_2/\mu_1'^2$ ratios are similar across the isochrone sets, there are large differences in $\Gamma_1$ and $\Gamma_2$ at some ages. This tells us, without even knowing the luminosity distribution function, that there are strong differences between both sets of isochrones, which can be directly related with differences in the treatment of stellar evolution, e.g. differences in the lifetimes of different phases.

\subsection{Application to the Virtual Observatory}

The method described in this work can be implemented in the VO as an automatized tool for the analysis of observed data, since synthetic models can be given in terms of probability distributions suitable to be used in data mining algorithms or in Bayesian analysis. To achieve this goal, an appropriate theoretical data model is necessary; the definition of such model is a task that is currently being carried out by the 
Theory interest group of the International Virtual Observatory Alliance ({\tt http://www.ivoa.net}). In addition, the extension of this 
probabilistic
formalism to distributions of luminosity ratios, which are used in diagnostic diagrams, would be an asset for the development of more robust VO analysis tools.

\section{Conclusions}\label{sect:conclusions}

This paper considers a series of problems in population synthesis that arise as a consequence of the distributed nature of stellar populations, and develops a new probabilistic formalism that takes them into account. 
With this formalism it is possible to reproduce and explain the features of Monte Carlo simulations without the need of performing them. The new formalism has several advantages with respect to Monte Carlo simulations in terms of generality and reliability. Unlike Monte Carlo simulations, it is not affected by sampling errors in the estimation of the moments. Unfortunately, 
its exact application requires computing repeated convolutions, and 
we do not know any computational tool that can perform this task efficiently.
Although the formalism is complete for luminosities, it must still be extended to the case of ratios. 
A summary of our conclusions follows:

\begin{itemize}

\item In the first part, we revise the current standard formalism and discuss
the phenomena it fails to address and the coincidences and differences with our method. The main difference between the standard approach and ours is that the former interprets the results of synthesis models deterministically, whereas our formalism accounts for their statistical distribution. We show how the resulting distribution functions can be characterized in terms of their moments and cumulants, and how the shape of such distributions can be obtained from data, information that is necessary to compute the confidence intervals of the models' predictions.

\item Standard synthesis models work by handling the 
sLDF and not the pLDF. Nonetheless, we show that pLDFs
are the convolution of 
sLDFs,
so that synthesis models can be used to study stellar populations, either integrated or resolved, once the correct 
probabilistic
formalism is included.

\item The cumulants of the distribution scale with the 
size
 of the stellar populations. We explain these scale relations and show how the cumulants can be obtained in terms of the distribution moments.
The variance as computed by synthesis models (generally in the context of SBF) is biased, as it actually corresponds to the second raw moment of the distribution, which is not a scalable quantity. Fortunately, given the high asymmetry and the power-law nature of the luminosity distribution, the numerical difference between the second raw and central moments is almost negligible. It is advisable, however, to use the right formula.

\item The standard deviation $\sigma$ cannot be used as the unit of confidence intervals unless the distribution
is known to be quasi-Gaussian.
A nearly Gaussian regime is reached only when the sample contains more than $\grsim 10^5$ stars, with the precise limit depending on the spectral region among other factors. It is therefore mandatory to perform safety checks before using tests that assume normality, such as the $\chi^2$ test. In particular, clusters with 
large skewness
should be treated with extreme caution,
since in those cases the bulk of the distribution
is several $\sigma$s away from the mean value. 
We refer to \cite{GGS04} for a more detailed analysis of this issue.

\item The customary assumption of a Poisson distribution in bins is, in fact, not accurate enough. A realistic solution, fully consistent with the underlying distributions, is the multinomial distribution. The multinomial distribution describes in a natural way the covariance effects introduced by the binning, i.e. the correlations between different bins.

\item We give a few guidelines for assessing the robustness of fits and show that not all the features of the electromagnetic spectrum are equally suitable for using in the fitting of theoretical to observational data. Although the relative dispersion $\sigma({\cal L})/M'_1$ tends to 0 for increasing $N_\mathrm{tot}$, $\sigma({\cal L})$ increases.

\item For practical applications, we show how the results of synthesis models can be directly applied to the study of resolved stellar populations in quite simple terms, i.e. obtaining the distribution moments from the observed stars. This formulation allows reliable comparisons of observed data with theoretical stellar models to be made.

\item Current synthesis models cannot be used for comparison with observed SBF, since an additional distribution function must be included in the treatment, describing the distribution of number of stars across pixels. 
Our formalism provides a natural way to do it.

\item We give a preliminary example of how the probabilistic formalism can be applied to the distribution of globular clusters and to chemical evolution models. The latter would require the inclusion of stochastic differential equations.

\item Finally, we derived implications of this study for 
the development of analysis tools 
in the VO.
For example,
we are implementing synthesis models that include the probabilistic formalism in a new VO tool called {\it PGos3}, under development at {\tt http://ov.inaoep.mx}. The present formalism will be also used for the development of data mining and in Bayesian algorithms. 

\end{itemize}

As our understanding of stellar populations shifts, population synthesis tools evolve. The problem of predicting the integrated properties of stellar populations
was initially framed as a deterministic one and solved by standard codes. A growing awareness of the spread in the input parameters has boosted the interest in Monte Carlo simulations, whose phenomenological exploration has brought about important insights into the statistical aspect of the problem. The time is ripe now for a further forward step, one that advances the problem from a statistical to a probabilistic formulation. As this evolution takes place, however, it is important to keep in mind that the new formalism reinterprets previous conceptions, rather than overthrowing them, and that it does not supersede the old tools, but instead aims to specify how and when they can be applied. The probabilistic formalism is best seen as a unifying model that includes the old tools and empowers them, in a direction that 
is becoming imperative for understanding the new observational data.

\begin{acknowledgements}
  This work has been the effort of many years of reflection on the sampling problems of the IMF. Therefore, there are lots of people that at some moment or other have contributed to this work with comments and suggestions. First of all, we would like to acknowledge Luis Manuel Sarro Bar\'o, whose incisive questions shed light on the problem of distributions in bins. We also acknowledge useful discussions with Marat Gilfanov. Steve Shore, Juan Betancort, Enrique P\'erez, and David Valls-Gabaud made useful suggestions on statistics, while Georges Meynet, Daniel Schaerer, Sandro Bressan, Scilla Degl'Innocenti, Enzo Brocato, and Leo Girardi helped us to understand track interpolation issues. We acknowledge Gabriella Raimondo, Rosa Amelia Gonz\'alez, and Gustavo Bruzual for several useful conversations on the subject of SBF and Monte Carlo simulations. We thank Alberto Buzzoni for very instructive discussions of synthesis models. The first draft of the paper was greatly improved thanks to the constructive criticism and encouraging comments of an anonymous referee. Joli Adams, our A\&A language editor, displayed infinite patience in helping us find just the right word. 
NCL acknowledges Carlos, Dario, and Eva for providing experimental facts showing that the application of 
mean values to individual data is a statistical fallacy.

This work was supported by the Spanish {\it Programa Nacional de Astronom\'\i a y Astrof\'\i sica} through the project AYA2004-02703. MC is supported by a {\it Ram\'on y Cajal} fellowship. VL is supported by a {\it CSIC-I3P} fellowship. 
\end{acknowledgements}

\appendix
\section{Notation}\label{app:notation}

This section contains an exhaustive summary of the notation used and its rationale.

\subsection{Quantities related to stars}

\begin{itemize}
 \item[$l_i$] is the luminosity of an individual object in a discrete sum (Eq. \ref{eq:Ltot_sum}).

\item[$\ell$] is a continuous variable representing the possible luminosity values of individual objects.

\item[$\varphi_L$] $= \varphi_L(\ell;t)$ is the probability density function (PDF) of the variable $\ell$, i.e. the stellar luminosity distribution function (sLDF) (Eq. \ref{eq:LDFtrad}). It depends on the assumed age ($t$, which is explicited here as a parameter), but also on the metallicity, the evolutionary tracks adopted, and the star formation history.

\item[$\phi_L$] $= \phi_L(p;t)$ is the characteristic function of $\varphi_L$, defined as its Fourier transform (Eq. \ref{eq:charfsLDF}). 

\item[$\mu'_n$] $= \mu'_n(t)$ is the n-th raw moment of the sLDF  (Eqs. \ref{eq:2nd_momentp}$-$\ref{eq:3-4th_moment}).

\item[$\mu_n$] $= \mu_n(t)$ is the n-th central moment of the sLDF  (Eqs. \ref{eq:2nd_moment}$-$\ref{eq:3-4th_moment}). The second central moment is the variance of the sLDF, and its square root, 
$\sigma$, is the standard deviation of the probability distribution. Note that in the literature the symbol $\sigma$ is often used to represent
the standard deviation of the sample, or mean square error, which is a statistical quantity. In this paper, we do not use $\sigma$ with this statistical meaning.

\item[$\kappa_n$]$ = \kappa_n(t)$ is the n-th cumulant of the sLDF (Eq. \ref{eq:kappadef}).

\item[$\gamma_n$] $= \gamma_n(t)$ are the skewness ($\gamma_1$)  and the kurtosis ($\gamma_2$) of the sLDF (Eqs. \ref{eq:gamma1}$-$\ref{eq:gamma2}).
\end{itemize}

\subsection{Quantities related to the integrated properties of a stellar population}

All these quantities depend on the total number of stars $N_\mathrm{tot}$ in the population, as well as on the age of the population:

\begin{itemize}

\item[$L_\mathrm{tot}$] is the integrated luminosity of a given individual population (Eq. \ref{eq:Ltot_sum}).

\item[${\cal L}$] $= {\cal L}(t)$ is a continuous variable representing the possible values of the integrated luminosity of stellar populations with $N_\mathrm{tot}$ stars (Eq. \ref{eq:Ltot_mean}).

\item[$\varphi_{\mathrm{L_{tot}}}$] $=\varphi_{\mathrm{L_{tot}}}({\cal L};t)$ is the probability density function of $\cal L$, i.e. the population luminosity distribution function (pLDF). Under suitable assumptions, it can be obtained by convolving the sLDF $N_\mathrm{tot}$ times with itself (Sect.~\ref{sub:exact_pLDF}).

\item[$\phi_\mathrm{L_{tot}}$] $= \phi_\mathrm{L_{tot}}(P;t)$ is the characteristic function of $\varphi_{\mathrm{L_{tot}}}({\cal L})$, defined as its Fourier transform (Eq. \ref{eq:char_pop}). 

\item[$M'_n$] $= M'_n(t)$ is the n-th raw moment of the pLDF.

\item[$M_n$] $= M_n(t)$ is the n-th central moment of the pLDF. If $n=2$, its square root is the variance of the pLDF, denoted as $\sigma({\cal L})$ in this paper.

\item[$K_n$] $= K_n(t)$ is the n-th cumulant of the pLDF. Under suitable assumptions, simple scale relations hold between $K_n$ and $\kappa_n$ (Eq. \ref{eq:scale}).

\item[$\Gamma_n$] $= \Gamma_n(t)$ is the skewness ($\Gamma_1$) and the kurtosis ($\Gamma_2$) of the pLDF. Under suitable assumptions, simple scale relations hold between $\Gamma_n$ and $\gamma_n$ (Eqs. \ref{eq:g1_G1}$-$\ref{eq:g2_G2}).

\end{itemize}

\subsection{Quantities related to the IMF}

\begin{itemize}
\item[$\varphi_M$] $ = \varphi_M(m)$ is the IMF, defined as the probability density function for a star of having an initial mass $m$ (Eq. \ref{eq:IMF1}).

\item[$\varphi'_M$] $ = \varphi'_M(m)$ is a function proportional to $\varphi_M(m)$:  $\varphi'_M(m) = \varphi_M(m) / \langle m \rangle$, where $\langle m \rangle$ is the mean mass of the IMF (Eq. \ref{eq:IMF2}).

\item[$w_i$] $= w_i(m^\mathrm{low}_i,m^\mathrm{up}_i;t)$ is the probability that a given star has a mass belonging to the interval $[m^\mathrm{low}_i,m^\mathrm{up}_i]$ (Eq.\ref{eq:wi}). Note that the interval is defined arbitrarily depending on the computational needs and the characteristic evolutionary time of the population considered, so in most cases it is varied depending on the age of the stellar population. 

\item[$n_i$] $= n_i(m^\mathrm{low}_i,m^\mathrm{up}_i;t)$ is a random (distributed) variable representing the number of stars in a given mass interval $[m^\mathrm{low}_i,m^\mathrm{up}_i]$ (Eqs. \ref{eq:varwiP}, \ref{eq:muwi_MC}). Its mean value is $\langle n_i \rangle = N_\mathrm{tot} \times w_i$.

\end{itemize}

\bibliographystyle{apj}

\begin{thebibliography}{}

\bibitem[Allar et al.(2001)]{2001ApJ...556..357A}Allard, F., Hauschildt, P., Alexander, D.r., Tamanai, A., and Schweitzer, A. 2001, ApJ, 556, 357

\bibitem[Bertelli et al.(1994)]{Be94}Bertelli, G., Bressan, A., Chiosi, C.,
   Fagotto, F., \&  Nasi E. 1994, \aaps , 106, 275 

\bibitem[Blinnikov \& Moessner(1998)]{BM98} Blinnikov, S., 
\& Moessner, R.\ 1998, \aaps, 130, 193 

\bibitem[Bressan et al.(1994)]{Breetal94} Bressan, A., Chiosi, 
C., \& Fagotto, F.\ 1994, \apjs, 94, 63 

\bibitem[Brocato et al.(1999)]{Broetal99} Brocato, E., 
Castellani, V., \& Romaniello, M.\ 1999, \aap, 345, 499 

\bibitem[Bruzual(2002)]{Brutuc}Bruzual, G. 2002,  in IAU Symp. 207,  
   Extragalactic star clusters, ed. D. Geisler, E. Grebel, and D. Minniti
   (San Francisco: Astr. Soc. Pacific), 616 (astro-ph/0110245)

\bibitem[Bruzual \& Charlot(2003)]{BC03} Bruzual, G., \& 
Charlot, S.\ 2003, \mnras, 344, 1000 

\bibitem[Buzzoni(1989)]{Buzz89}Buzzoni, A. 1989, \apjs , 71, 871

\bibitem[Cantiello et al.(2003)]{CRBC03} Cantiello, M., Raimondo, G., 
Brocato, E., \& Capaccioli, M.\ 2003, \aj, 125, 2783 

\bibitem[Cariulo et al.(2004)]{Caretal04} Cariulo, P., 
Degl'Innocenti, S., \& Castellani, V.\ 2004, \aap, 421, 1121 

\bibitem[Castellani et al.(2003)]{Casetal03} Castellani, V., 
Degl'Innocenti, S., Marconi, M., Prada Moroni, P.~G., \& Sestito, P.\ 2003, 
\aap, 404, 645 

\bibitem[Castelli and Kurucz(2003)]{2003IAUS..210..A20C}Castelli, F. and Kurucz, R. L. 2003 in IAU Symp. 210 {\it Modeling of Stellar Atmospheres} Eds. N. Piskunov, W.W. Weis and D.F. Gray, PosterA20 ({\tt http://kurucz.harvard.edu/})

\bibitem[Cervi{\~ n}o \& Luridiana(2004)]{CL04} Cervi{\~ 
n}o, M., \& Luridiana, V.\ 2004, \aap, 413, 145 

\bibitem[Cervi{\~ n}o \& Luridiana(2005)]{CL05} Cervi{\~ 
n}o, M., \& Luridiana, V.\ 2005, in {\it Resolved Stellar Populations}, Eds. D. Valls-Gabaud and M. Ch\'avez, ASP Conf. Ser., in press (astro-ph/0510411) 
 
\bibitem[Cervi{\~n}o \& Mas-Hesse(1994)]{CMH94} 
   Cervi{\~n}o, M.,~\& Mas-Hesse, J.~M.\ 1994, \aap, 284, 749 

\bibitem[Cervi\~no \& Valls-Gabaud(2003)]{CVG03}Cervi\~no, M., \&  
   Valls-Gabaud, D. 2003, \mnras, 338, 481

\bibitem[Cervi{\~n}o, Luridiana, \& Castander(2000)]{CLC00}
   Cervi{\~n}o, M., Luridiana, V., \& Castander, F.~J.\ 2000, \aap, 360, L5 

 \bibitem[Cervi{\~n}o et al.(2001a)]{Cetal01} Cervi{\~n}o, M., 
  G{\'o}mez-Flechoso, M.~A., Castander, F.~J., Schaerer, D., Moll{\'a}, M., 
  Kn{\"o}dlseder, J., \& Luridiana, V.\ 2001a, \aap, 376, 422 

\bibitem[Cervi{\~ n}o et al.(2000)]{Cgam00} Cervi{\~ n}o, M., 
Kn{\" o}dlseder, J., Schaerer, D., von Ballmoos, P., \& Meynet, G.\ 2000, 
\aap, 363, 970 

\bibitem[Cervi{\~ n}o et al.(2003)]{Cetal03} Cervi{\~ n}o, M., 
Luridiana, V., P{\' e}rez, E., V{\'{\i}}lchez, J.~M., \& Valls-Gabaud, D.\ 
2003, \aap, 407, 177 

\bibitem[Cervi\~no et al.(2002b)]{CVGLMH02}Cervi\~no, M.,  
   Valls-Gabaud, D., Luridiana, V., \& Mas-Hesse, J.M. 2002b, \aap, 381, 51

\bibitem[Charlot \& Bruzual(1991)]{CB91} Charlot, S., \& 
Bruzual, A.~G.\ 1991, \apj, 367, 126 

\bibitem[Chiosi et al.(1988)]{Chietal88} Chiosi, C., Bertelli, 
G., \& Bressan, A.\ 1988, \aap, 196, 84

\bibitem[Cid Fernandes et al.(2005)]{CFetal05} Cid Fernandes, R., 
Mateus, A., Sodr{\' e}, L., Stasi{\' n}ska, G., \& Gomes, J.~M.\ 2005, 
\mnras, 358, 363 

\bibitem[Gilfanov et al.(2004)]{GGS04} Gilfanov, M., Grimm, 
H.-J., \& Sunyaev, R.\ 2004, \mnras, 351, 1365  

\bibitem[Girardi(2000)]{Gproc00}Girardi, L. 2000, 
   in ASP Conf. Ser. 221, Massive
   Stellar Clusters, eds. A. Lan{\c{c}}on, and C. Boily,
   (San Francisco: Astr. Soc. Pacific), 133 (astro-ph/0001434)

\bibitem[Girardi(2002)]{Gituc}Girardi, L. 2002,  in IAU Symp. 207,  
   Extragalactic star clusters, eds. D. Geisler and E. Grebel (San  
   Francisco: Astr. Soc. Pacific), 625 (astro-ph/0108198)

\bibitem[Girardi and Bertelli(1998)]{GB98} Girardi, L.~and Bertelli, G.\ 1998, \mnras, 300, 533

\bibitem[Girardi et al.(2000)]{Gi00} Girardi, L., Bressan, A., Bertelli,
  G., \& Chiosi, C. 2000, \aaps, 141, 317
  
\bibitem[Girardi et al.(2002)]{Gi02} Girardi, L., Bertelli, 
  G., Bressan, A., Chiosi, C., Groenewegen, M.~A.~T., 
  Marigo, P., Salasnich, B., \& Weiss, A.\ 2002, \aap, 391, 195

\bibitem[Gonz{\' a}lez et al.(2004)]{GLB04} Gonz{\' a}lez, 
R.~A., Liu, M.~C., \& Bruzual A., G.\ 2004, \apj, 611, 270 

\bibitem[Gonz{\'a}lez Delgado et al.(2005)]{GDetal04}Gonz{\'a}lez Delgado, R.M., Cervi{\~n}o, M., Martins, L.P., Leitherer, C., Hauschildt, P.H. 2005, \mnras 357, 945 ({\tt http://www.iaa.es/$\sim$rosa/HRES} [ascii] and {\tt http://ov.inaoep.mx/} [VO Table]) 

\bibitem[Gray and Corbally(1994)]{1994AJ....107..742G} Gray, R.O., Corbally, C.J., 1994, AJ, 107, 742 ({\tt http://www1.appstate.edu/dept/physics/spectrum/spectrum.html})

\bibitem[Hauschildt and Baron(1999)]{1999JCAP..102...41H}Hauschildt, P., and Baron, E. 1999, Journal of Computational and Applied Mathematics, 102, 41

\bibitem[Hubeny, Lanz, and Jeffery(1995)]{HLJ95} Hubeny, I., Lanz, T., Jeffery, C.S., 1995, SYNSPEC - A User's Guide

\bibitem[Jamet et al.(2004)]{Jametal04} Jamet, L., P{\' e}rez, 
E., Cervi{\~ n}o, M., Stasi{\' n}ska, G., Gonz{\' a}lez Delgado, R.~M., \& 
V{\'{\i}}lchez, J.~M.\ 2004, \aap, 426, 399 

\bibitem[Kendall \& Stuart(1977)]{KS77}Kendall, M., \& Stuart, A. 1977, The advanced theory of statistics, vol. 1 (London: Griffin)

\bibitem[Kroupa(2001)]{Kro01} Kroupa, P.\ 2001, \mnras, 322,  231 

\bibitem[Kurth et al.(1999)]{KFAF99} Kurth, O.~M., Fritze-v.~Alvensleben, U., \& Fricke, K.~J.\ 1999, \aaps, 138, 19

\bibitem[Kurucz(1991)]{1991sabc.conf..441K}Kuruck, R.L. 1991, in {\it Stellar Atmospheres, Beyond Clasical Limits}, Crivellari L., Hubeny I., Hummer D.G. eds., Kluwer, Dordrecht, p. 441 (ATLAS9) ({\tt http://kurucz.harvard.edu/})

\bibitem[Lanz and Hubeny(2003)]{2003ApJS..146..417L}Lanz, T. and Hubeny, I. 2003, ApJ 146, 417 ({\tt http://tlusty.gsfc.nasa.gov/})

\bibitem[Leitherer et al.(1999)]{SB99} Leitherer, C.~et al.\ 
  1999, \apjs, 123, 3 

\bibitem[Lucy(2000)]{L00} Lucy, L.~B.\ 2000, \mnras, 318, 92 

\bibitem[Maraston(1998)]{Mar98} Maraston, C.\ 1998, \mnras, 
300, 872 

\bibitem[Marigo \& Girardi(2001)]{MG01}Marigo, P., \& Girardi, L. 2001, 
   \aap, 377, 132 
   
\bibitem[Martins et al.(2005)]{Metal04}Martins, L.P., Gonz{\'a}lez Delgado, R.M., Leitherer, C., Cervi{\~n}o, M., and Hauschildt, P.H. 2005, \mnras, 358, 49

\bibitem[Mas-Hesse \& Kunth(1991)]{MHK91} Mas-Hesse, J.~M., 
\& Kunth, D.\ 1991, \aaps, 88, 399 

\bibitem[Miller \& Scalo(1979)]{MS79} Miller, G.~E., \& 
Scalo, J.~M.\ 1979, \apjs, 41, 513 

\bibitem[Pellerin(2005)]{P05} Pellerin, A.\ 2005, \aj  ~(in press, astro-ph/0510686)

\bibitem[Renzini \& Buzzoni(1986)]{RB86} Renzini, A., \& 
Buzzoni, A.\ 1986, ASSL Vol.~122: Spectral Evolution of Galaxies, 195 

\bibitem[Salpeter(1955)]{Sal55} Salpeter, E. E. 1955, \apj, ~121, 161

\bibitem[Santos \& Frogel(1997)]{SF97}Santos, J. F. C. Jr., \& 
   Frogel, J.A. 1997, \apj, 479, 764 

\bibitem[Schaerer et al.(1993a)]{Schetal93} Schaerer, D., Charbonnel, C., Meynet, G., Maeder, A., \& Schaller, G.\ 1993a, \aaps, 102, 339

\bibitem[Schaller et al.(1992)]{Schetal92}Schaller, G., Schaerer, D.,  Meynet, G. \& Maeder, A.\ 1992, \aaps, 96, 269

\bibitem[Tout et al.(1996)]{Touetal96} Tout, C.~A., Pols, O.~R., 
Eggleton, P.~P., \& Han, Z.\ 1996, \mnras, 281, 257 

\bibitem[Wilson \& Hurley(2003)]{WH03} Wilson, R.~E., \& Hurley, J.~R.\ 2003, \mnras, 344, 1175

\bibitem[Zhang \& Fall(1999)]{ZF99} Zhang, Q., \& Fall, 
S.~M.\ 1999, \apjl, 527, L81 


\end{thebibliography}
\end{document}